\documentclass[rmp,twocolumn,floatfix,epsfig,graphics]{revtex4} 

\usepackage{amsmath,amsfonts,amssymb,graphics,graphicx,epsfig,times,color,bbm,fancybox}

\usepackage[sort&compress]{natbib}

\newcommand{\dist}{\text{dist}}
\newcommand{\me}{\mathrm{e}}
\newcommand{\mi}{\mathrm{i} }
\newcommand{\md}{\mathrm{d}}

\newtheorem{corollary}{Corollary}
\newtheorem{theorem}{Theorem}

\newtheorem{lemma}{Lemma}
\newtheorem{conjecture}{Conjecture}

\newsavebox{\oszifour}

\begin{document}

\title{Area laws for the entanglement entropy -- a review}

\author{J.\ Eisert,$^{1,2,3}$ M.\ Cramer,$^{3,4}$ and M.B.\ Plenio$^{3,4}$}

\affiliation{$^1\,$Institute of Physics and Astronomy,
University of Potsdam, 14469 Potsdam, Germany\smallskip\newline
$^2\,$Institute for Advanced Study Berlin,
14193 Berlin, Germany\smallskip\newline
$^3\,$Institute for Mathematical Sciences, Imperial College 
London, London SW7 2PG, UK\smallskip\newline
$^4\,$Institut f{\"u}r Theoretische Physik, University of Ulm, 89069 Ulm, Germany}
\begin{abstract}
Physical interactions in quantum many-body systems are typically
local: Individual constituents interact mainly with their few 
nearest neighbors. This locality of interactions is inherited 
by a decay of correlation functions, but also reflected by scaling laws of a 
quite profound quantity: The entanglement entropy of ground 
states. This entropy of the reduced state of a subregion often 
merely grows like the boundary area of the subregion, and not 
like its volume, in sharp contrast with an expected extensive 
behavior. Such ``area laws'' for the entanglement entropy and 
related quantities have received considerable attention in 
recent years. They emerge in several seemingly unrelated fields, 
in the context of black hole physics, quantum information science, 
and quantum many-body physics where they have important 
implications on the numerical simulation of lattice models. 

In this Colloquium we review the current status of area laws 
in these fields. Center stage is taken by rigorous results 
on lattice models in one and higher spatial dimensions. The 
differences and similarities between bosonic and fermionic 
models are stressed, area laws are related to the velocity 
of information propagation in quantum lattice models, and 
disordered systems, non-equilibrium situations, and topological 
entanglement entropies are discussed. These questions are 
considered in classical and quantum systems, in their ground 
and thermal states, for a variety of correlation measures. A 
significant proportion of the article is devoted to the clear 
and quantitative connection between the entanglement content 
of states and the possibility of their efficient numerical 
simulation. We discuss matrix-product states, higher-dimensional 
analogues, and variational sets from entanglement renormalization 
and conclude by highlighting the implications of area laws on 
quantifying the effective degrees of freedom that need to be 
considered in simulations of quantum states.
\end{abstract}

\maketitle

\newcommand{\R}{\mathbbm{R}}
\newcommand{\rr}{\mathbbm{R}}
\newcommand{\nn}{\mathbbm{N}}
\newcommand{\cc}{\mathbbm{C}}
\newcommand{\zz}{\mathbbm{Z}}
\newcommand{\id}{\mathbbm{1}}
\newcommand{\ee}{\mathbbm{E}}

\newcommand{\tr}{{\rm tr}\,}
\newcommand{\gr}[1]{\boldsymbol{#1}}
\newcommand{\ket}[1]{|#1\rangle}
\newcommand{\bra}[1]{\langle#1|}
\newcommand{\avr}[1]{\langle#1\rangle}
\newcommand{\G}{{\cal G}}
\newcommand{\eq}[1]{eq.~(\ref{#1})}
\newcommand{\ineq}[1]{Ineq.~(\ref{#1})}
\newcommand{\sirsection}[1]{\section{\large \sf \textbf{#1}}}
\newcommand{\sirsubsection}[1]{\subsection{\normalsize \sf \textbf{#1}}}

\newcommand{\proofend}{\hfill\fbox\\\medskip }

\tableofcontents
  
\section{Introduction}

In classical physics concepts of entropy quantify the extent 
to which we are uncertain about the exact state of a physical 
system at hand or, in other words, the amount of information
that is lacking to identify the microstate of a system from 
all possibilities compatible with the macrostate of the system.
If we are not quite sure what microstate of a system to expect,
notions of entropy will reflect this lack of knowledge. 
Randomness, after all, is always and necessarily related to 
ignorance about the state.

\begin{figure}[hbt]
\includegraphics[width=0.8\columnwidth]{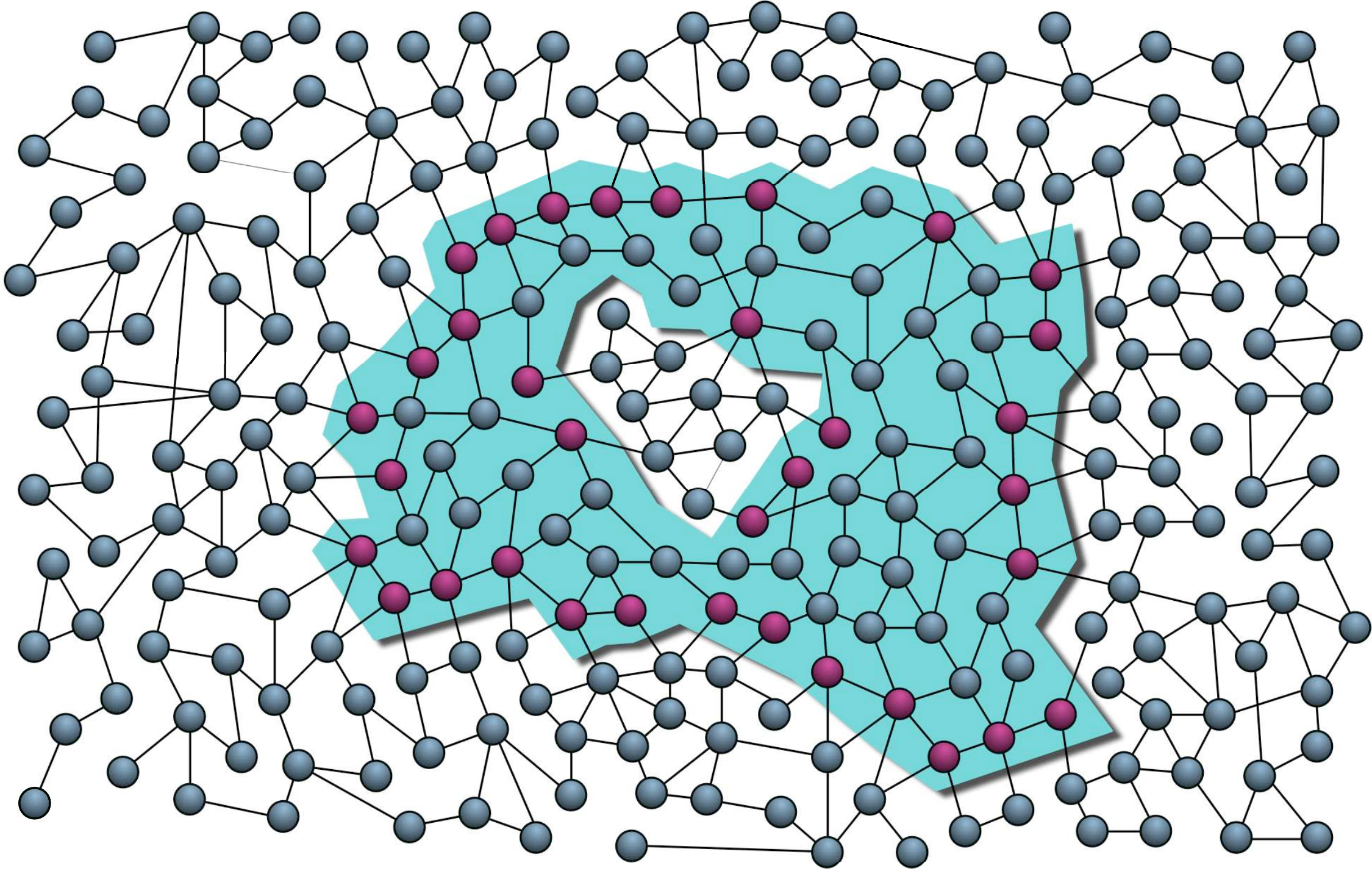}
\caption{\label{areaBoebbel}A lattice $L$ with a distinguished set $I\subset L$ (shaded area). Vertices
depict the boundary $\partial I$ of $I$ with surface area $s(I)=|\partial I|$.}
\end{figure}

In quantum mechanics positive entropies may arise even without an 
objective lack of information. To see this, let us consider a 
quantum lattice systems (see e.g.,\ Fig.\ 1) as an example for a 
quantum many-body system where each of the vertices $i$ of 
the lattice $L$ is associated with an individual quantum system. 
This quantum many-body system is thought to be in its non-degenerate 
pure ground state $\rho=|\psi\rangle\langle\psi|$ at zero temperature
which has vanishing {\it von-Neumann entropy} 
\begin{equation*}
        S(\rho) = - \text{tr}[\rho \log_2 \rho].
\end{equation*}
Let us now distinguish a region of this quantum lattice 
system, denoting its sites with the set $I$ and all other 
sites with $O=L\backslash I$. If we consider the reduced state 
$\rho_I= \text{tr}_{O}[\rho]$ of the sites of the region $I$, 
the state will not be pure in general and will have a 
non-vanishing von-Neumann entropy $S(\rho_I)$.
\footnote{Of interest are also other entropies, such as the 
{\it Renyi entropies}, $S_\alpha(\rho) =(1-\alpha)^{-1} 
\log_2 \text{tr} [\rho^\alpha]$ with $\alpha\geq 0$. For 
$\alpha\searrow 1$ the usual von-Neumann entropy is recovered. 
In particular in the context of simulatability, Renyi entropies 
for arbitrary $\alpha$ play an important role.}

In contrast to thermal states this entropy does not originate 
from a lack of knowledge about the microstate of the system. 
Even at zero temperature we will encounter a non-zero 
entropy! This entropy arises because of a very fundamental property 
of quantum mechanics: Entanglement. This quite intriguing 
trait of quantum mechanics gives rise to correlations even
in situations where the randomness cannot be traced back 
to a mere lack of knowledge. The mentioned quantity, the entropy 
of a subregion is called {\it entanglement entropy} or  
{\it geometric entropy} and, in quantum information,
{\it entropy of entanglement}, which 
 represents an operationally defined entanglement 
measure for pure states (for recent reviews see refs.\ \cite{Horodecki,InHouseReview}).

In the context of quantum field theory, questions of scaling
of entanglement entropies  in the size of $I$ have some tradition. Seminal work on the 
geometric entropy of the free Klein-Gordon field 
\cite{Bombelli,Srednicki} and subsequent work on conformal field
theories \cite{Larsen,Callan,PreskillOld,Calabrese,HardStuff} was 
driven in part by the intriguing suggested connection to the 
Bekenstein-Hawking black hole entropy 
\cite{Bekenstein,Hawking74,BekensteinCP}. 

In recent years, studies of properties of the entanglement entropy 
in this sense have enjoyed a revival initiated in refs.\ \cite{Harmonic,OldTobias,OldFazio,Latorre1}.
Importantly, this renewed activity is benefitting from the new 
perspectives and ideas of quantum information theory, and from 
the realisation of their significance for the understanding of 
numerical methods and especially their efficiency for describing
quantum many-body 
physics. Quantum information theory also provides novel conceptual 
and mathematical techniques for determining properties of the 
geometric entropy analytically.
 
At the heart of these studies are questions like: What role do 
genuine quantum correlations---entanglement---play in quantum many-body systems? 
Typically, in such investigations, one abstracts to a large extent
from the microscopic specifics of the system: Quite in the spirit 
of studies of {\it critical phenomena}, one thinks less of very 
detailed properties, but is rather interested 
in the {\it scaling} of the entanglement entropy when the distinguished 
region grows in size.  In fact, for 
quantum chains, this scaling of entanglement as genuine quantum 
correlations---a priori very different from the scaling of two-point 
correlation functions---reflects to a large extent the critical 
behavior of the quantum many-body system, and shares some
relationship to conformal charges.

At first sight one might be tempted to think that the entropy of
a distinguished region $I$, will always possess an extensive 
character. Such a behavior is referred to as a {\it volume 
scaling} and is observed for thermal states. Intriguingly, for 
typical ground states, however, this is not at all what one 
encounters: Instead, one typically finds an {\it area law}, or 
an area law with a small (often logarithmic) correction: This 
means that if one distinguishes a region, the scaling of the 
entropy is merely linear in the {\it boundary area} of the 
region. The entanglement entropy is then said to fulfill an 
{\it area law}. It is the purpose 
of this article to review studies on area laws and the scaling 
of the entanglement entropy in a non-technical manner. 

The main four motivations to approach this question (known to 
the authors) are as follows:

\begin{itemize}

\item {\bf The holographic principle and black hole entropy:} 
The historical motivation to study the entanglement or geometric 
entropy stems from considerations of black hole physics: It 
has been suggested in the seminal work of refs.\ 
\cite{Bombelli,Srednicki} that the area law of the geometric 
entropy for a discrete version of a massless free scalar 
field---then numerically found for an imaginary sphere in 
a radial symmetry---could be related to the physics of 
{\it black holes},\cite{HardStuff} in particular the 
Bekenstein-Hawking entropy of a black hole which is proportional
to its boundary surface. It has been muted that 
the {\it holographic principle} \cite{Holographic}---the 
conjecture that the information contained in a volume of 
space can be represented by a theory which lives in the 
boundary of that region---could be related to the area 
law behavior of the entanglement entropy in microscopic
theories.

\item {\bf Distribution of quantum correlations in quantum 
many-body systems:} Area laws also say something quite profound 
on how quantum correlations are distributed in ground states of 
local quantum many-body systems. Interactions in quantum many-body 
systems are typically local, which means that systems interact 
only over a short distance with a finite number of neighbors. 
The emergence of an area law then provides support for the
intuition that short ranged interactions require that quantum 
correlations between a distinguished region and its exterior 
are established via its boundary surface. That a strict area 
law emerges is by no means obvious from the decay of two-point 
correlators, as we will see. Quantum phase transitions are 
governed by quantum fluctuations at zero temperature, so it 
is more than plausible to observe signatures of {\it criticality} 
on the level of entanglement and {\it quantum correlations}. This 
situation is now particularly clear in one-dimensional systems, 
\cite{Harmonic,Latorre1,Fannes,Korepin,FranciniLong,Calabrese,Latorre2,Single,Its,Keating,Farkas,OldBriegel,Quench,Vidal,JVidalLong,Vidal2,Casini,Cardy2,OneD,Cardy3,Review}
but progress has also been made in higher-dimensional systems,\cite{Area,Area2,Area3,Graphs,Quench2,LatorreReview,PEPS,Wolf,FarkasLong,Preskill,Fradkin}
with rigorous area laws specifically for quasi-free bosonic \cite{Area,Area2,Area3} and 
fermionic \cite{Wolf,Klich,Halfspace,FarkasLong} systems,
as well as in
disordered systems \cite{Refael2}.

\item {\bf Complexity of quantum many-body systems and their 
simulation:} One of the key motivations for 
studying area laws stems from a quite practical context: The 
numerical simulation of quantum many-body systems. In fact, if 
there is little entanglement in a ground state of a many-body
systems, one might suspect on intuitive grounds that one can 
describe this ground state with relatively few parameters. More 
specifically, for one-dimensional systems, 
one would expect numerical algorithms like the powerful 
{\it density-matrix renormalization group method} \cite{White,Scholl} 
(DMRG) to perform well if the ground state contains a small 
amount of entanglement. This suspicion can in fact be made 
rigorous \cite{Peschel,DMRG,OneD,SchuchApprox} as it turns out 
that the scaling of entanglement specifies how well a given state 
can be approximated by a matrix-product state \cite{FCS,Scholl} 
as generated in DMRG. It is hence not the decay behavior of 
correlation functions as such that matters here, but in fact 
the scaling of entanglement.

\item {\bf Topological entanglement entropy:} 
The topological entanglement entropy is an indicator of 
{\it topological order},\cite{OldWen,WittenOld,Ortiz} 
a new kind of order in quantum 
many-body systems that cannot be described by local order 
parameters \cite{Preskill,Wen,FradkinTopological,Haque,Ortiz}. 
Lattice models having a non-vanishing topological entanglement 
entropy may be seen as lattice instances of topological 
quantum field theories. Here a global feature is detected by means
of the scaling of geometric entropies.
\end{itemize}

In this Colloquium we do not have sufficient space to give 
an account of all known derivations of area laws for the 
entanglement entropy. However, we will try 
not to merely remain at a superficial level and only state results, 
but will explain 
a number of key techniques and arguments. When we label main 
statements as ``theorems'' this is done to highlight their 
special role, to make it easier to follow the line of reasoning.
For details of arguments 
and proofs, often technically involved, we refer the reader to the 
original work. The reason for the technicality of proofs originates 
from the type of question that is posed: To distinguish a region 
of a lattice breaks the translational symmetry of the problem -- even in a translationally invariant setting.
While numerical studies are sometimes easier to come by, analytical 
argument can be technically involved, even for quasi-free 
models. In this article, we discuss the study of entanglement 
entropy primarily (i) from the viewpoint of quantum information 
theory, (ii) with an emphasis on rigorous and analytical results, 
and (iii) the implications on the efficiency of numerical
simulation.

\section{Local Hamiltonians and area laws}
Throughout this article, we will consider quantum many-body systems 
on a lattice. Such quantum lattice systems are ubiquitous in the 
condensed matter context \cite{Vojta} where they play a key role 
in obtaining an understanding of material properties from a 
microscopic basis. Lattices systems are also of considerable 
importance in the study of quantum field theories where a lattice
provides a natural ultra-violet cut-off and facilitates numerical
simulations of quantum fields \cite{LatticeFields}.
One could think, e.g., of systems of strongly correlated electron 
systems or lattice vibrations of a crystal lattice. 
With the advent of research on {\it cold atoms} in optical lattices, 
quantum lattice systems can also be prepared in laboratory conditions 
with an unprecedented degree of control \cite{BlochDZ08}.

We will consider---at least in parts of this article---general 
lattices. Each vertex of the {\it lattice} is associated with a 
quantum system, such as a spin, a bosonic or a fermionic system. 
It is convenient to think of this lattice as a simple graph 
$G=(L,E)$ with vertices $L$, and the edge set $E$ labeling 
neighborhood relations. $G$ could be the graph representing 
a one-dimensional chain with periodic boundary conditions, and 
in fact a good proportion of this article will deal with such 
quantum chains. For later purposes, it will be convenient to 
think in terms of such a slightly more general picture, however. 
The Hilbert space of the total many-body 
system is then the tensor product
        \begin{equation*}
        {\cal H}= \bigotimes_{j\in L} {\cal H}_j
         \end{equation*}
where ${\cal H}_j$ is the Hilbert space associated with the physical
system on lattice site $j$. On such a lattice, one has $\dist(j,k)$ 
for $j,k\in L$ as the natural {\it graph theoretical distance}
which is the length of the shortest path connecting $j$ and $k$.
For a {\it cubic lattice} of dimension ${\cal D}$ with periodic 
boundary conditions, in turn, $\dist({j},{k})= \sum_{d=1}^{\cal D} 
|j_d - k_d|$, where the components of $j,k\in L$ are taken modulo 
the base length of the cubic lattice.

We will be concerned largely with {\it local}
Hamiltonians on lattices. This means that the physical system
associated with a specific lattice site will interact only with
its neighbors and {\em not} with all sites of the lattice.
The total Hamiltonian can hence be written as
\begin{equation*}
        H = \sum_{X\subset L} H_X,
\end{equation*}  
where $H_X$ has a compact support $X$, independent of the system size,
that is the number of lattice sites denoted by $|L|$. 

The {\it boundary surface area} $s(I)$ of a distinguished region 
$I$ of the lattice $L$ can be defined in a very natural fashion 
on such a graph as the cardinality of the set of boundary points
\begin{equation}\label{BoundaryArea}
\partial I=\left\{ i\in I:
\,\text{there is a } j\in L\backslash I \text{ with } dist(i,j)=1\right\},
\end{equation}
so $s(I)=|\partial I|$, see Fig.\ \ref{areaBoebbel}. Throughout the 
article, unless defined specifically otherwise, we will say that the
entanglement entropy satisfies an {\it area law} if
\begin{equation*}
        S(\rho_I) = O(s(I)).
\end{equation*}
This means that the entropy of the reduced state $\rho_I$ scales 
at most as the boundary area of the region $I$.

Before we dive into the details of known results on area laws in 
quantum many-body systems, let us appreciate how unusual it is
for a quantum state to satisfy an area law.
In fact, a quantum state picked at random will exhibit a very 
different scaling behavior. If one has a lattice system with 
$d$-dimensional constituents and divides it into a subsystem 
$I\subset L$ and the complement $O=L\backslash I$, then one 
may consider the expected entanglement entropy of $I$ for the 
natural choice, the unitarily invariant Haar measure. One finds 
\cite{Page,RandomEnt,SenRandom}
\begin{equation*}
        \ee[S(\rho_I)] > |I|\log_2(d)- \frac{d^{|I|-|O|}}{2 \log_2(2)}.
\end{equation*}
That is, asymptotically, the typical entropy of a subsystem
is almost maximal, and hence linear in the number of constituents
$|I|$.
Hence a ``typical'' quantum state will asymptotically satisfy a 
volume law, and not an area law. As we will see that area laws 
are common for ground states of quantum many-body systems, we 
find that in this sense, ground states are very non-generic. 
This fact is heavily exploited in numerical 
approaches to study ground states of strongly correlated 
many-body systems: One does not have to vary over all
quantum states in variational approaches, but merely
over a much smaller set of states that are good candidates
of approximating ground states of local Hamiltonians well,
that is states that satisfy an area law.

\section{One-dimensional systems}

Most known results on area laws refer to one-dimensional chains
such as {\it harmonic} or {\it spin chains}. This emphasis is 
no surprise: After all, a number of 
physical ideas---like the Jordan-Wigner transformation---as well 
as mathematical methods---such as the theory of Toeplitz determinants 
and Fisher-Hartwig techniques---are specifically tailored to 
one-dimensional translationally invariant systems.

If we distinguish a contiguous set of quantum systems of a chain,
a {\it block} $I=\{1,\dots, n\}$ the boundary of the block consists 
of only one (two) site(s) for open (periodic) boundary conditions.
An area law then clearly means that the entropy is upper bounded 
by a constant independent of the block size $n$ and the lattice 
size $|L|$, i.e., 
\begin{equation}\label{Area}
        S(\rho_I)= O(1).
\end{equation}
We will see that in quantum chains, a very clear picture emerges 
concerning the scaling of the entanglement entropy. Whether an 
area law holds or not, will largely depend on whether the system 
is at a {\it quantum critical point} or not. We will summarize 
what is known in one-dimensional systems at the end of the 
detailed discussion of quantum chains, starting with bosonic 
harmonic chains.

\subsection{Bosonic harmonic chain}
Bosonic harmonic quantum systems, as well as fermionic models and 
their quantum spin chain counterparts like the XY model, play a 
seminal role in the study of quantum many-body systems. Harmonic 
lattice systems model discrete versions of Klein-Gordon fields,
vibrational modes of crystal lattices or of trapped ions and
serve generally as lowest order approximations to anharmonic 
systems. The fact 
that they are integrable renders even sophisticated questions 
like the scaling of the geometric entropy in instances amenable 
to fully analytical study, even in higher spatial dimensions. 
In fact, in the latter case these so-called quasi-free models 
are the only settings that allow for rigorous analytical results
so far. Hence, they do form the central object of consideration 
to explore what should be expected concerning general scaling laws.

The Hamiltonian for a harmonic lattice $L$ is given by
\begin{equation}
        \label{ham}
        {H}=\frac{1}{2}\sum_{i,j \in L} \bigl(
        p_i P_{i,j} p_j +  x_i X_{i,j} x_j \bigr),
\end{equation}
where $X,P \in \rr^{|L|\times |L|}$ are real, symmetric and positive 
matrices determining the coupling structure of the systems. The
canonical operators $x_i,p_i$ satisfy the canonical commutation 
relations $[x_j,p_k]=\mi\delta_{j,k}$. 
In terms of the bosonic 
annihilation operators $b_j = (x_j + \mi p_j)/\sqrt{2}$
the Hamiltonian eq.\ (\ref{ham}) reads
\begin{equation}
        \label{BosonHamilton}
        H=\frac{1}{2}\sum_{i,j}\bigl(
        b_i^\dagger A_{i,j}b_j+b_iA_{i,j}b_j^\dagger+b_iB_{i,j}b_j+b_i^\dagger B_{i,j}b_j^\dagger\bigr),
\end{equation}
where $A=(X+P)/2$, $B=(X-P)/2$.
Ground and thermal states of the above Hamiltonian are fully 
characterized by the second moments of the canonical 
operators, while first moments vanish \cite{EisertP03} (entanglement properties
of the state are invariant under changes of first moments anyway). The second moments define the 
{\it covariance matrix} 
\begin{eqnarray}
        \Gamma_{i,j} =\langle  \left\{r_i, r_j\right\}\rangle
        =\langle r_i r_j\rangle+\langle r_j r_i\rangle,
\end{eqnarray}
where ${{r}}=({x}_1,\dots,{x}_{|L|},{p}_1,\dots,{p}_{|L|})$ is 
the vector of canonical operators. The covariance matrix of the
ground state of eq.\ (\ref{ham}) is given by 
$\Gamma=\Gamma_x\oplus\Gamma_p$, 
where 
\begin{equation*}
	\Gamma_p =  X^{1/2} \bigl(X^{1/2} P X^{1/2}\bigr)^{-1/2} X^{1/2} 
\end{equation*}
and $\Gamma_x=\Gamma_p^{-1}$, see refs.\ \cite{Area2,Schuch}. On
the level of covariance matrices unitary operations express themselves
as symplectic transformations $S$ that preserve the commutation
relations $\sigma_{k,l} = \mi [r_k,r_l]$, i.e., $S\sigma S^T=\sigma$. Importantly, 
{\it Williamson's Theorem} states that for any strictly 
positive matrix $A\in \rr^{2N\times 2N}$ there exist a symplectic 
transformation $S$ such that $SAS^T = D$, where $D$ is a diagonal 
matrix with the same spectrum as the positive square roots of $(\mi \sigma A)^{2}$. The eigenvalues 
$d_i$ of $D$ are called the {\it symplectic eigenvalues} of $A$. 

Now, what is the entanglement content of the ground state? To
answer this we need to define entanglement measures and compute 
them in terms of the properties of the covariance matrix. The first 
of these is of course the entropy of entanglement.
Williamson's theorem shows that any function of a state that is 
unitarily invariant is fully determined by the symplectic 
eigenvalues. Notably, the entropy of a Gaussian state $\rho$ 
with symplectic eigenvalues $d_1,\dots, d_N$ of the covariance matrix of $\rho$ is given by
\begin{equation*}
        S(\rho) =\sum_{j=1}^N \biggl(
        \frac{d_j+1}{2} \log_2 \frac{d_j+1}{2}
        - 
        \frac{d_j-1}{2}\log_2\frac{d_j-1}{2}
        \biggr).
\end{equation*}

A key ingredient in the analytical work is another full entanglement
measure that was defined in quantum information theory, the 
{\it logarithmic negativity} \cite{PhD,Volume,VidalNegativity,PlenioNegativity,InHouseReview,EisertPlenioneg}.
It is defined as  
\begin{equation*}
        E_{N}(\rho,I) = \log_2 \|\rho^{\Gamma_I}\|_1,
\end{equation*}
where $\|A\|_1=\text{tr}[(A^\dagger A)^{1/2}]$ is the trace norm
and $\rho^{\Gamma_I}$ is the {\it partial transpose} of $\rho$ with 
respect to the interior $I$. The partial transpose w.r.t. the 
second subsystem is defined as 
$(|i\rangle \langle k|\otimes|j\rangle \langle l|)^{\Gamma_2}=
|i\rangle \langle k|\otimes|l\rangle \langle j|$ and extended by linearity.
On the level of covariance matrices the partial transpose is 
partial time reversal, i.e. $p_i\mapsto -p_i$ if $i\in I$ while 
$x_i$ remains invariant. Then for $\rho$ with covariance matrix 
$\Gamma=\Gamma_x\oplus\Gamma_p$ we find that $\rho^{\Gamma_{I}}$ 
has covariance matrix $\Gamma^\prime = \Gamma_x\oplus 
(F\Gamma_pF)$, where the diagonal matrix 
$F$ has entries $F_{i,j}=\pm \delta_{i,j}$, depending on whether a coordinate is in 
$I$ or $O$:
Then one finds for a state with covariance matrix $\Gamma=\Gamma_x\oplus\Gamma_p$ the logarithmic negativity 
\cite{Harmonic,marcusThesis}
\begin{equation*}
        E_{N}(\rho,I) = \frac{1}{2}\sum_{k=1}^{|L|} \log_2\max \bigl\{1, \lambda_k\bigl(\Gamma_p^{-1}F\Gamma_x^{-1}F\bigr)\bigr\},
\end{equation*}
where the $\{\lambda_k\}$ denote the eigenvalues.
The logarithmic negativity has two key features. Mathematically, 
the importance of $E_{N}(\rho,I)$ is due to 
\begin{equation}
        E_{N}(\rho,I) \geq S(\rho_I)\label{logengineq}
\end{equation}
which holds for all pure states $\rho$. This {\it upper bound} 
for the entanglement entropy is simpler to compute as one does 
not have to look at spectra of reductions $\rho_I$ but of the 
full system. This renders a study of area laws possible even in 
higher dimensional systems. Secondly, in contrast to the entropy
of entanglement, the negativity is also an entanglement measure 
for mixed states, such as thermal states and provides an upper
bound on other important measures of mixed state entanglement \cite{VedralP98,Bennett,ChristandlW04,InHouseReview}.

All of the above holds for general lattices $L$ but for the moment 
we will focus on the one-dimensional setting, that is 
$L=\{1,\dots, N\}$ where $N$ is even to allow us to consider 
the {\it symmetrically bisected chain} $I=\{1,\dots, N/2\}$ with 
periodic boundary conditions and $P=\id$. We concentrate on
the ground state and discuss thermal states
 later.\footnote{We do not discuss the entanglement
properties in excited states here as this area has not been 
explored in detail so far \cite{Excited,Buzek04}.} It is worth noting
that in higher spatial dimension the natural analog of this
setting, the half-space, is of some importance as it allows 
for a reduction of the problem in question to the 1-D case 
discussed here \cite{Halfspace}. Furthermore, the scaling 
behavior of the entanglement of the half-chain has 
direct consequences on the availability of efficient 
representations of the state by means of matrix-product states
as will be discussed in some detail later on in this article.
For a general nearest-neighbor 
coupling this means that $X$ is the circulant matrix, 
\begin{equation}\label{HC}
        X=\text{circ}(a,b,0,\dots, 0,b),
\end{equation}
as a consequence of translational invariance. $b$ 
specifies the coupling strength, $a$ defines the on-site
term, $\lambda_{\text{min}}(X)=a-2|b|$, i.e., positivity
demands $a>2|b|$, and the energy gap above the ground state
is given by  $\Delta E = \lambda^{1/2}_{\text{min}}(XP ) = 
(a-2|b|)^{1/2}$. For the logarithmic negativity of the 
symmetrically bisected half-chain we find \cite{Harmonic}:
\begin{theorem}[Exact negativity of the half-chain]
\label{T1}
        Consider a Hamiltonian of a harmonic chain on 
        $L=\{1,\dots, N\}$ with
        periodic boundary conditions, $P=\id$, and nearest-neighbor 
        interactions as in eq.\  (\ref{HC}).
        Then the entanglement entropy of the symmetrically
        bisected chain and the logarithmic negativity satisfy
        \begin{equation}\label{halfent}
                S(\rho_I)\leq E_{N}(\rho,I) = \frac{1}{4}\log_2 
                \left(
                \frac{a+2|b|}{a-2|b|}
                \right)
                = \frac{1}{2}\log_2 
                \left(
                \frac{ \|X\|^{1/2}}{\Delta E}
                \right)
        \end{equation}
where $\|\cdot \|$ is the operator norm and 
$\Delta E = \lambda^{1/2}_{\text{min}}(X)$.      
\end{theorem}

The quantity $\|X\|$ will later be related to the speed of sound 
in the system. This expression for the block entanglement quantified 
with respect to the negativity is exact and no approximation. This 
was to the knowledge of the authors a first rigorous area law for 
a lattice system, complementing earlier seminal work for fields \cite{Callan}. 
Remarkably, this expression is entirely independent 
of $N$, the system size. The most important observation here is 
that an area law holds, which can be expressed in terms of the
spectral gap in the system: Whenever the system is 
non-critical in the sense that the energy gap $\Delta E$ satisfies
$\Delta E \ge c > 0$ with a system size independent constant $c$, 
a one-dimensional area law will hold.
The above link of entanglement entropy and spectral
gap in the system can be established in much more generality 
and we will delay this discussion to a later subsection. 

The argument leading to Theorem \ref{T1} is involved, and for 
details we refer to ref.\ \cite{Harmonic}. The  interesting 
aspect of this proof is that the spectrum of the half chain 
can not be obtained analytically, thus not allowing for a 
direct computation of the entanglement content. Instead, it 
is the particular combination of spectral values of the 
partial transpose entering $E_{N}(\rho,I)$ itself that can be 
explicitly computed. The proof makes heavy use of the symmetry 
of the problem, namely the invariance under a flip of the two 
half chains. 

This result suggested that the locality of the interaction
in the gapped model is inherited by the locality of
entanglement, a picture that was also later confirmed
in more generality. Note that the above bound is a
particularly tight one, and that it may well suggest what 
prefactor in terms of the energy gap and speed of sound
one might expect in general area laws, as we will discuss
later.

Let us now consider an important model for which the energy 
gap vanishes in the thermodynamical limit $N\rightarrow\infty$:
Taking $a=m^2+2N^2$, $b=-N^2$, identifying lattice sites by 
$i=xN$, and the canonical operators by $x_i=N^{-1/2}\phi(x)$, 
$p_i=N^{-1/2}\pi(x)$, one obtains the Klein-Gordon field 
Hamiltonian
\begin{equation}
\label{KG}
        {H}=\frac{1}{2}\int_{0}^1\!\!\md x\,\left(\pi^2(x)
        +\left(\frac{\partial}{\partial x}\phi(x)\right)^2
        +m^2\phi^2(x)\right),
\end{equation}
in the field limit $N\rightarrow\infty$. 
(For a detailed discussion of the continuum limit for the Klein-Gordon
field, see also ref.\ \cite{Botero}.)  From the expression 
(\ref{halfent}) for the entanglement, we immediately obtain
\begin{equation}
\label{KGLogNeg}
        E_{N}(\rho,I)=\frac{1}{4}\log_2\left(1+\frac{4N^2}{m^2}\right)
        \rightarrow_{N\rightarrow\infty} \frac{1}{2}\log_2 \left(\frac{2N}{m}\right).
\end{equation}
This is a striking difference to the 
area laws that we have 
observed earlier, now the entanglement does not saturate but
diverges with the length of the half-chain.\footnote{Compare also 
the divergence of the entanglement entropy in collectively
interacting chains \cite{Fleischhauer}.} The behavior 
observed here will be mirrored by a similar logarithmic 
divergence in critical quantum spin chains and fermionic
systems. This will be discussed in the following section.

\subsection{Fermionic chain and the XY model} 

Following the initial work on bosonic models of ref.\ 
\cite{Harmonic} similar questions were explored in 
fermionic systems and the associated spin models.
The numerical studies in refs.\ \cite{Latorre1,Latorre2} 
presented a significant first step in this direction. 
Their key observation, later confirmed rigorously 
\cite{Korepin,Its,Keating} using techniques that we will 
sketch in this section, is that the scaling of the 
entanglement entropy as a function of the block size 
appears to be related to the system being {\it quantum critical 
or not}. Again, for a gapped system, away from a quantum 
critical point, the entanglement entropy would saturate, 
i.e., an area law holds. In turn in all cases when the system 
was critical, the numerical study indicated that the 
entanglement entropy grows beyond all bounds. More 
specifically, it grows logarithmically with the block 
size. This behavior is also consistent with the behavior of geometric entropies 
in conformal field theory \cite{Callan,Larsen} which applies to the critical
points of the models discussed in 
refs.\ \cite{Latorre1,Latorre2,CriticalKorepin,Franchini}. 
The intriguing aspect here is that being critical 
or not is not only reflected by the scaling of expectation 
values of two-point correlators, but in fact by the ground 
state entanglement, so genuine quantum correlations.

This section defines the setting, introduces the basic 
concepts required and outlines the rigorous results in
more detail. Fermionic quasi-free models, that is Hamiltonians 
that are quadratic in fermionic operators $f_i$ and $f_i^{\dagger}$, 
\begin{equation}\label{GeneralFermionic}
        H= \frac{1}{2}\sum_{i,j\in L}
        \left(
        f_i^\dagger A_{i,j} f_j
        - 
        f_i A_{i,j} f_j^\dagger
        +f_i B_{i,j} f_j
        -f_i^\dagger B_{i,j} f_j^\dagger
        \right)
\end{equation}
may be treated by similar analytical techniques and follow similar
intuition to the bosonic case. 
In eq.\ (\ref{GeneralFermionic}), to ensure Hermiticity of the
Hamiltonian, $A^T=A$ and $B^T=-B$ must hold for the matrices $A$ 
and $B$ defining the coupling. The role of the canonical coordinates
is taken by the {\it Majorana operators} $x_j = (f_j^\dagger+ f_j)/
\sqrt{2}$ and $p_j = \mi (f_j^\dagger - f_j)/\sqrt{2}$,
while the role of symplectic transformations is taken by 
orthogonal transformations. The energy gap above the ground state is given by
the smallest non-zero singular value of $A+B$. 

Note that, in contrast to the bosonic case, the ground state 
is $2^{|L|-\text{rank}(A+B)}=:q$-fold degenerate. We define
the ground state expectation $\langle\cdot\rangle=\text{tr}
[\cdot P_0]/q$, where $P_0$ projects onto the ground state 
sector. Then, as in the bosonic case, the ground state is 
fully characterized by two-point correlations embodied in 
the covariance matrix with entries
\begin{equation*}
        -\mi\Gamma_{i,j} = \langle\left[r_i,r_j\right]\rangle
        = \langle r_ir_j\rangle-\langle r_jr_i\rangle,
\end{equation*}
where now $r=(x_1,\dots,x_{|L|},p_1,\dots,p_{|L|})$ collects 
Majorana operators. One then finds
\begin{equation}\label{FermionicGamma}
        \Gamma=\left(\begin{array}{cc}
           0&-V\\
           V^T&0
        \end{array}\right),\;\;\; V=|A+B|^{+}(A+B),
\end{equation} 
where $\cdot^+$ indicates the Moore-Penrose generalized inverse 
of a matrix \cite{HornJohnson}, i.e., for a unique ground state one 
simply has $V=|A+B|^{-1}(A+B)$. The entropy of a contiguous
block $I$ of fermions in the ground state can be expressed 
in terms of the singular values $\sigma_k$ of the principle submatrix 
$V_I$ of $V$. One finds $S(\rho_I)=\sum_{k}f(\sigma_k)$, 
where
\begin{equation}
\label{fermionEntropy}
        f(x)=-\frac{1-x}{2}\log_2\left(\frac{1-x}{2}\right)-
        \frac{1+x}{2}\log_2\left(\frac{1+x}{2}\right).
\end{equation}
All the above holds for general lattices but for the moment
we will turn to a discussion of $L=\{1,\dots, N\}$.

We have started the discussion on the level of fermionic
operators to highlight the similarity to the bosonic case. It
is important to note however, that these fermionic models share 
a close relationship to natural spin models in the 1-D setting.
This is revealed by the {\it Jordan Wigner transformation} which 
relates fermionic operators with spin operators according to
\begin{eqnarray}
        \sigma^z_i= 1- 2 f_i^\dagger f_i,\;\;
        \frac{\sigma^x_i+ \mi \sigma^y_i}{2} = \prod_{k=1}^{i-1} (1-2 f_k^\dagger f_k) f_i,
\end{eqnarray}
where $\sigma^x_i,\sigma^y_i,\sigma^z_i$ denote the Pauli 
operators associated with site $i\in L$. The fermionic model 
eq.\ (\ref{GeneralFermionic}) is hence equivalent to a spin model 
with short or long-range interactions.

The most important model of this kind is the XY model with a 
transverse magnetic field, with nearest-neighbor interaction, 
$A_{i,i}= \lambda$, $A_{i,j}= -1/2$ if $\text{dist}(i,j)=1$,
and $B_{i,j}=- B_{j,i}= \gamma/2$ for $\text{dist}(i,j)=1$.
This gives rise to 
\begin{equation}\label{xyh}
        H=-\frac{1}{2}\sum_{\langle i,j\rangle}
        \left(
        \frac{1+\gamma}{4}\sigma^x_i \sigma^x_j + 
        \frac{1-\gamma}{4}\sigma^y_i\sigma^y_j
        \right) - \frac{\lambda}{2}\sum_{i\in L}
        \sigma^z_i,
\end{equation}
where $\langle i,j\rangle$ denotes summation over nearest neighbors,
$\gamma$ is the anisotropy parameter, and $\lambda$ an external 
magnetic field.\footnote{Note that the boundary conditions give
rise to a (sometimes overlooked) subtlety here. For open 
boundary conditions in the fermionic model, the Jordan Wigner 
transformation relates the above fermionic model to the spin 
model in eq.\ (\ref{xyh}) with open boundary conditions. For 
periodic boundary conditions, the term $f^\dagger_N f_1$ is 
replaced by the operator $(\prod_{j=1}^N (2f_j^\dagger f_j-1) )
f^\dagger_N f_1$. Hence, strictly speaking, the periodic 
fermionic model does not truly correspond to the periodic 
XY model \cite{XYCircle}. Importantly, the degeneracy of 
ground states is affected by this. For a degenerate ground 
state, the {\it entanglement of formation} \cite{Bennett}, 
the {\em relative entropy of entanglement} \cite{VedralP98}, 
the {\it distillable entanglement} \cite{Bennett} or the 
logarithmic negativity of the ground state 
sector are the appropriate entanglement 
quantifiers \cite{Horodecki,InHouseReview}, and no 
longer the entropy of entanglement. Only in the 
case that for large subsystems $n$ one can almost 
certainly locally distinguish the
finitely many different ground states, the entropy
of entanglement for each of the degenerate 
ground states still gives the correct value for the entanglement of a subsystem.} 
Once again, translational invariance of the model 
means that the spectrum can be readily computed by means of
a discrete Fourier transform. One obtains
\begin{equation*}
        E_k = \left({
        (\lambda- \cos(2\pi k/N))^2 + \gamma^2 \sin^2(2\pi k/N)
        }\right)^{1/2},
\end{equation*}
for $k=1,\dots, N$. This is a well-known integrable
model \cite{LSM,BC}.

In the plane defined by $(\gamma,\lambda)$ several critical 
lines can be identified: Along the lines $|\lambda |=1$ and 
on the line segment $\gamma=0$, $|\lambda|\le 1$, the system 
is critical, $\lim_{N\rightarrow\infty} \Delta E(N) =0$. 
For all other points in the $(\gamma,\lambda)$ plane the 
exists a $c>0$ independent of $N$ such that $\Delta E\ge c$.  
The class of models with $\gamma=1$ are called {\it Ising model}.
The most important case subsequently is the isotropic case
of the XY model, then often referred to as {\it XX model} 
or isotropic XY model. This is the case when $\gamma=0$. 
The XX model is critical whenever $|\lambda|\le 1$. 
The XX model is
equivalent to the {\it Bose-Hubbard model} in the 
limit of hard-core bosons, so the Bose-Hubbard model
with the additional constraint that each site can be occupied
by at most a single boson.

Let us assume that we have a non-degenerate
ground state, such that the entropy of entanglement
$S(\rho_I)$ really quantifies the entanglement content.
For the translation-invariant system at hand, the 
 entries $V_{i,j}=V_{i-j}$ of $V$ are given by 
 \begin{equation}
 \label{XYcirculant}
         V_{l}=\frac{1}{|L|} \sum_{k=1}^{|L|} 
        g_k e^{2\pi \mi  lk/{|L|}},\;\;\;
        g_k=\lambda_k\left(\frac{A+B}{|A+B|}\right).
 \end{equation}
The entanglement properties of the model are encoded in 
the numbers $g_k$. For $N=|L|\rightarrow\infty$, we can
write
\begin{equation*}
         V_{l}=\frac{1}{ 2\pi} \int_0^{2\pi} \md\phi\,
        g(\phi) e^{\mi l\phi},
\end{equation*}
for $| l|\leq N/2$, where $g:[0,2\pi)\rightarrow \cc$ is 
called the {\it symbol} of $ V$. Note that the {\it Fermi 
surface} is defined by the discontinuities of the symbol. In order to evaluate 
the entropy of a reduction $S(\rho_I)$, we merely
have to know the singular values of $n\times n$-submatrices 
$V_I=V_n$ of $V$, see eq.\ (\ref{fermionEntropy}).
For isotropic models,
i.e., for $B=0$, $V$ then being symmetric, the singular 
values are the absolute values of the eigenvalues. 
In other words,  in order to understand the correlation and entanglement
structure of sub-blocks of such systems, one has to 
understand properties of matrices the entries of which are
of the form $T_{i,j}=T_{i-j}$. Such matrices are called 
{\it Toeplitz matrices}. An $n\times n$ Toeplitz matrix is 
entirely defined by the $2n-1$ numbers $T_l$, $l=1-n,\dots,n-1$. 

The spectral values $\lambda_1( T_n),\dots, \lambda_n(T_n)$
of $ T_n$ are just the zeros of the characteristic polynomial 
\begin{equation*}
        \text{det}( T_n-\lambda\id) = \prod_{k=1}^n
        (\lambda_k( T_n) - \lambda),
\end{equation*}  
so in order to grasp the asymptotic behavior of the spectrum 
of $T_n$, it is sufficient to know the asymptotic behavior 
of this determinant expression.\footnote{Once this quantity
is known, one can evaluate the
entropy by means of 
\begin{eqnarray}
        S(\rho_I)=\sum_{k=1}^n f(|\lambda_k|)= \lim_{\varepsilon\rightarrow 0}
        \oint
        \frac{\md z}{2\pi \mi} f_\varepsilon( z)\frac{\partial}{\partial z}
        \log \text{det}( V_n-z\id).
\end{eqnarray}
Here, the integration path in the complex plane has been chosen 
to contain all eigenvalues  $\lambda_k( V_n)$. The function $ f_\varepsilon$ 
is a continuation of $f$: We require that 
$\lim_{\varepsilon\rightarrow 0}f_\varepsilon(z)=f(|z|)$, 
including the parameter $\varepsilon$ such that  $f_\varepsilon$ 
is analytic within the contour of integration.}
The mathematical theory of determinants of such {\it Toeplitz
matrices} is very much developed. The
Fisher-Hartwig Theorem provides exactly the tools
to study the asymptotic behavior of Toeplitz determinants
in terms of the symbol. Crudely speaking, what matters are
the zeros of $g$ and the jumps: Once $g$ is written in 
what can be called a normal form, one can ``read off'' the asymptotic
behavior of the sequence of Toeplitz determinants
defined by this symbol. Note that the matrices $ V_n-z\id$
take here the role of $T_n$. The exact formulation of the
Fisher-Hartwig Theorem is presented in the appendix.

This machinery was used in refs.\ \cite{Korepin,Its,Keating,Single} to evaluate the asymptotic
behavior of the block entropy for the critical XX model and 
other isotropic models. In the first paper introducing this
idea \cite{Korepin}, in fact, there is a single jump from 
$1$ to $-1$ in the symbol defining the Toeplitz matrices 
(and no zeros), which gives rise to the prefactor of $1/3$ of 
$\log_2 (n)$ in the formula for the entanglement entropy in 
the XX model. This prefactor---which emerges here rather 
as a consequence of mathematical properties of the symbol---is
related to the conformal charge of the underlying conformal 
field theory. In more general isotropic models, as has been 
pointed out in ref.\ \cite{Keating}, the number of jumps 
determines the prefactor in the entanglement scaling. 
Hence in such quasi-free isotropic fermionic models,
the connection between criticality and a logarithmic 
divergence is very transparent and clear:
If there is no Fermi surface at all, and hence no jump in 
the symbol, the system will be gapped and hence non-critical.
Then, the entropy will saturate to a constant. 

In contrast, in case there is a Fermi surface, this will 
lead to jumps in the symbol, and the system is critical.
In any such case one will find a logarithmically divergent
entanglement entropy. The prefactor is determined by the number
of jumps. So more physically speaking, what matters is the
number of boundary points of the Fermi surface in the interval 
$[0,2\pi)$. So---if one can say so in a simple one-dimensional
system---the ``topology of the Fermi surface determines the 
prefactor''. This aspect will later be discussed in more 
detail. Refs.\ \cite{Korepin,Its,Keating} find the following:

\begin{theorem}[Critical quasi-free fermionic
chains]
\label{CriticalLogarithmic}
Consider a family of quasi-free isotropic fermionic 
Hamiltonians with periodic boundary conditions as in eq.\ 
(\ref{GeneralFermionic}) with $B=0$. Then, the entanglement 
entropy of a block of $I=\{1,\dots, n\}$ continuous spins 
scales as
\begin{equation*}
        S(\rho_I) = \xi \log_2(n) + O(1).
\end{equation*}
$\xi >0$ is a constant that can be related to the number of 
jumps in the symbol  (defined above). This applies, e.g., 
to the scaling of the entanglement entropy in the XX spin
model, for which
\begin{equation*}
        S(\rho_I) = \frac{1}{3} \log_2(n) + O(1).
\end{equation*}
\end{theorem}

The constant $\xi$ is not to be mistaken for the conformal
charge which will be discussed later.
These arguments correspond to the isotropic model with $B=0$, 
where the Fisher-Hartwig machinery can be conveniently applied. 
In contrast, the anisotropic case, albeit innocent-looking, 
is overburdened with technicalities. Then, in order to 
compute the singular values of submatrices of $V$ as in 
eq.\ (\ref{FermionicGamma}), it is no longer sufficient to 
consider Toeplitz matrices, but {\it  block Toeplitz} matrices 
where the entries are conceived as $2\times 2$ matrices. 
This setting has been studied in detail in ref.\ \cite{Its} 
in case of a non-critical anisotropic system, finding again 
a saturation of the entanglement entropy and in ref.\ 
\cite{BlockToeplitz} where the prefactor of the area law 
for the entanglement entropy in the gapped XX model was
computed rigorously. 
Ref. \cite{Franchini} discusses also other Renyi entropies in 
this model.

Using an idea that originates from the concept of single-copy 
entanglement all these technicalities may be avoided and we 
can prove that the entanglement entropy diverges at least 
logarithmically in case of a critical (anisotropic) Ising 
model. The $\Omega$ notation just means that there is 
asymptotically a lower bound with
this behavior.\footnote{$f(n)=\Omega(g(n))$ if
$\exists C>0,n_0:\forall n>n_0: |Cg(n)|\leq |f(n)|$.} 

\begin{theorem}[Divergence
for the critical Ising model]
The entanglement entropy in the critical Ising model scales as
\begin{equation}\label{SCB}
        S(\rho_I) = \Omega(\log_2 n).
\end{equation}
\end{theorem}

The starting point leading to this result from 
ref.\  \cite{Single} is a lower bound in the operator
norm of $\rho_I$ leading to
\begin{eqnarray}
        -\log_2 ||\rho_I||_{\infty}&=& -\log_2 \text{det}((\id +  V_n)/2)\nonumber\\
        &\geq&  -\frac{1}{2}\log_2 |\text{det}( V_n)|.
\end{eqnarray}
This makes a big difference: We now no longer need the 
singular values of $V_n$ (which would lead to an enormously 
complicated block Toeplitz expression, for a case for which 
the Fisher-Hartwig conjecture has not yet been proven).
Instead---as the absolute value of the determinant is just 
as well the product of the absolute values of the eigenvalues 
as of the singular values---we can use the ordinary 
Fisher-Hartwig machinery to get an asymptotic handle on 
eigenvalues. For the critical Ising model, we can again 
find an explicit factorization of the Fisher-Hartwig-symbol,
in terms of a function reflecting a single discontinuity and
an analytical function. Using again a proven instance of the 
Fisher-Hartwig conjecture \cite{Libby}---albeit a different 
one than used in the case of an isotropic model---one finds 
the bound as in eq.\ (\ref{SCB}); for details see
ref.\ \cite{Single}.  The entanglement in two blocks of the
critical Ising model has been studied in ref.\ 
\cite{Pascazio}.

Another useful starting point to obtain bounds to entanglement
entropies in fermionic systems is to make use of quadratic 
bounds to the entropy function: Such quadratic bounds 
immediately translate to a bound to the function $f$ in the 
expression of the entropy of a fermionic state in terms of 
the covariance matrix as in eq.\ (\ref{fermionEntropy}), as
\begin{equation}\label{fbound}
        ({1-x^2})^{1/2}\geq f(x)\geq 1-x^2.
\end{equation}
This immediately translates to a bound of the
form $\text{tr}[(\id - V_IV_I^T)^{1/2}]\geq S(\rho_I)\geq 
\text{tr}[\id - V_IV_I^T]$, where
$V_I$ is the submatrix of $V$
associated with the interior $I$ \cite{Fannes}. These bounds
have also been exploited in the higher-dimensional analysis in
ref.\ \cite{Wolf}.

A method to obtain area laws in particular for symmetrically 
bisected quantum 
chains is the so-called method of {\it corner transfer matrices}.
This method has first been used in ref.\ \cite{Calabrese} 
for the computation of the entanglement entropy, 
using ideas going back to ref.\ 
\cite{PeschelEntanglementDMRG}. 
The infinite sum of ref.\ \cite{Calabrese} could be 
performed in ref.\ \cite{Peschel}, giving also rise to a formula for 
the entanglement entropy in the XX model. This idea has also
applied to further models in ref.\ \cite{Weston}.

To conclude the discussion of critical quasi-free fermionic
models let us note that the correspondence of being critical
(gapped) and having a logarithmically divergent (saturating)
entanglement entropy holds true for {\it local} systems only. 
If one allows for long-ranged interactions, then one can 
indeed find gapped, non-critical models that exhibit a 
logarithmically divergent entanglement entropy \cite{Quench}:
\begin{theorem}[Gapped model
with long-range interactions]
There exist models with long-range interactions, the 
coupling strength being bounded by $r/\text{dist}(j,k)$ 
for some constant $r>0$, such that for some
constant $\xi>0$
\begin{equation*}
        S(\rho_I) = \xi \log_2(n) + O(1).
\end{equation*}
\end{theorem}
Hence, being gapped---albeit having power-law correlations---does 
not necessarily imply an area law. If one 
allows for long-range interactions (and a {\it fractal 
structure} of the Fermi-surface), one can show that one 
can even approach arbitrarily well a {\it 
volume law} for the entanglement entropy \cite{Fannes,Farkas}.  
Interestingly, states that are defined by {\it quantum 
expanders} can have 
exponentially decaying correlations and still have large
entanglement, as has been 
proven in refs.\ \cite{HastingsExpander,Shma}. These models give
again 
rise to long-range Hamiltonians, but they still very clearly demonstrate
a strong distinction between correlations and entanglement.

\subsection{General gapped local spin models}

We will now turn to the discussion of general 1-D gapped spin 
models with local interactions, where each site supports a
$d$-dimensional quantum system. As it is stated rigorously
in the theorem below for such models {\it an area law always 
holds} \cite{OneD}. The proof is deeply rooted in the existence
of {\it Lieb-Robinson bounds} which have also been essential
in the proof of the exponential decay of correlation functions 
in gapped local models \cite{Decay,Decay2}.\footnote{This
result is compatible with an earlier result of an area
law in {\it 1-D gapped quantum field theories}, based on
the c-theorem presented in ref.\ \cite{Calabrese}. 
This work also connected
the role of the boundary points between 
regions $I$ and $O$ with the cluster decomposition in 
quantum field theory. In gapless system with open boundaries, the
entropy is then half of the one in the situation of 
having periodic boundary conditions.}

As we allow for arbitrary $d$, it is sufficient to consider
Hamiltonians on the chain $L=\{1,\dots, N\}$ that have 
interactions only to nearest neighbors. Then
\begin{equation}\label{NN}
        H=\sum_{j\in L} H_{j,j+1}
\end{equation}  
where $H_{j,j+1}$ is supported on sites $j$ and $j+1$. We also
impose a constraint of {\it finite-interaction strength}
in that the operator norm  $\| H_{i,i+1}\| \leq J$
for some $J>0$. Then ref.\ \cite{OneD} finds:

\begin{theorem}[Area law for gapped spin chains]
Consider a local Hamiltonian $H$ as in eq.\ (\ref{NN}) with 
finite interaction strength. Suppose $H$ has a unique ground 
state with a spectral gap $\Delta E$ to the first excited state. 
Let us as before consider the block $I=\{1,\dots, n\}$. Then, 
\begin{equation}
\label{entbd}
        S(\rho_I)\leq S_{\text{max}} = c_0 \xi\ln(6\xi) 
        \ln(d) 2^{6\xi \ln(d)},
\end{equation}
for some numerical constant $c_0>0$ of order unity, and where 
$\xi=\max (2v/\Delta E,\xi_C)$, $v$ is the velocity of sound 
and $\xi_C>0$ is of the order of unity.
\end{theorem}

The proof of this statement is quite intricate \cite{OneD}
and well beyond the scope in this article. 
At its heart is the way locality enters by virtue of the 
Lieb-Robinson Theorem. It is a statement on the existence 
of a {\it speed of sound} in local Hamiltonian systems with 
finite-dimensional constituents: Let us imagine we single 
out two disjoint sets $X,Y$ from a lattice, and consider 
observables $A$ and $B$ that have support only on $X$ and 
$Y$, respectively. Then $[A,B]=0$. If we evolve $A$ with 
time under a local Hamiltonian $H$ it is no longer exactly 
true that $A(t)$ and $B$ commute: $A(t)$ will be significantly
supported on more and more sites, ``melting away'', and 
developing a long tail in support. For short times or
large distances between sets $X$ and $Y$, the commutator 
of $A(t)$ and $B$ will be very small. How small exactly 
is governed by the {\it Lieb-Robinson Theorem} \cite{LR,OneD,Decay,Decay2}:

\begin{theorem}[Lieb-Robinson-Theorem]
\label{LRT}
Let $H$ be  as in eq.\ (\ref{NN}) a local Hamiltonian with
a finite interaction strength. Then there exists a velocity 
of sound $v>0$ and $\mu, c>0$ such that for any two 
operators $A$ and $B$ with support on disjoint sets $X$ and 
$Y$ we have that
\begin{equation}
        \label{lrb}
        \| [A(t),B]\| \leq
        c  \, \| A \|\,\, \| B  \|
        \exp\left(-\mu ({\text{dist}}(X,Y)-v|t| )\right),
\end{equation}
where the distance between sets is taken to be
${\text{dist}}(X,Y)=\min_{i\in X,j\in Y}(|i-j|)$, 
and where 
\begin{equation*}
        A(t)=e^{ \mi H t} A e^{-\mi H t}.
\end{equation*}
The velocity $v$ is of order $J$.
\end{theorem}

This statement, natural as it may seem when viewed with 
a reasonable physical intuition, is a rigorous, and profound 
statement on how locality manifests itself in quantum 
lattice systems. From this bound, the {\it decay of 
correlation functions} in gapped models can be proven \cite{Decay,Decay2},
an area law as above,\cite{OneD}, 
the quantization of the Hall conducance for interacting electrons \cite{Michalakis},
as well as statements concerning {\it propagation of 
quantum information and correlations} through local 
dynamics \cite{Quench2}.\footnote{The assumption that
we have a spin system, meaning finite-dimensional local 
constituents, is crucial here.} Lieb-Robinson bounds 
also feature in the proof of a higher-dimensional 
Lieb-Schultz-Mattis theorem \cite{LSM2,LSM3}.

We will later, in Subsection \ref{noneq}, encounter 
another consequence 
of the Lieb-Robinson theorem, namely that quenched 
non-equilibrium systems generically satisfy area laws when
starting from a product state and undergoing time evolution under
a local Hamiltonian. This perspective receives a  lot of attention
in the context of non-equilibrium dynamics of quantum many-body
systems. Here, the Lieb-Robinson is also the basis for the functioning
of numerical  {\it light cone methods} to study time evolution of 
quantum many-body systems {LightCone,MPS,NonMPS,NonMPS2},
in which effectively, only the essential part inside the causal cone 
is simulated.

\subsection{Results from conformal field theory}

In critical models the correlation length diverges and the 
models become scale invariant and allow for a description 
in terms of conformal field theories. 
According to the universality hypothesis, the microscopic 
details become irrelevant for a number of key properties.
These {\it universal} quantities then depend only on basic 
properties like the symmetry of the system, or the spatial 
dimension. Models from the same universality class are 
characterized by the same fixed point Hamiltonian under 
renormalization transformations, which is invariant under 
general rotations. Conformal field theory 
then describes such continuum models, which have the symmetry 
of the {\it conformal group} (including translations, rotations, 
and scalings). The universality class is characterized by the 
{\it central charge} $c$, a quantity that roughly quantifies 
the ``degrees of freedom of the theory''. For free bosons $c=1$, 
whereas the Ising universality class has $c=1/2$.

Once a model is known to be described by a conformal field theory,
powerful methods are available to compute universal properties, and
entanglement entropies (or even the full reduced spectra) of 
subsystems.\footnote{Conformal field theory provides---in this 
context specifically in $1+1$-dimensions---a powerful repertory 
of methods to compute quantities that are otherwise inaccessible
especially for non-integrable models. From a mathematical physics 
perspective, it is the lack of a rigorous proof of the relationship 
between the lattice model and the conformal field theory that
makes such a treatment, pedantically  speaking, non-rigorous.}
This approach applies for $1+1$-dimensional systems, that is
with one spatial dimension. In the seminal work ref.\ \cite{Larsen}
the entanglement entropy in $1+1$-dimensions has been calculated, see also 
refs.\ \cite{Callan,PreskillOld}. The work refs.\ \cite{Calabrese,Cardy3} put this
into a more general context, and also allow for non-contiguous 
regions $I$. The local spectra of the reductions are discussed 
in refs.\ \cite{Orus,OrusSpectrum,CalabreseSpectrum}. 
Block-block entanglement is also discussed in ref.\
\cite{Capelle,MarcovitchRPR08}. For a short non-technical review, see ref.\ \cite{Cardy2}. 

Starting point of the computations is the observation that powers 
of the reduced density matrix $\rho_I^n$ for any positive integer 
$n$ can be computed. The series 
$\text{tr}[\rho_I^n]= \sum_j \lambda_j(\rho_I)^n$ 
is absolute convergent and analytic for all
$\text{Re}(n)>1$. The derivative exists, and hence one can make use of
\begin{equation*}
        S(\rho_I)=\lim_{n\searrow 1}
        -\frac{1}{\ln 2} \frac{\partial}{\partial n}\text{tr}[\rho_I^n]
\end{equation*}
to compute the entanglement entropy. This procedure is 
typically referred to as {\it ``replica trick''}.
This leads in $1+1$-dimensions to the expression \cite{Larsen}
\begin{equation}
        \label{confent}
        S(\rho_I) = \frac{c}{3}\log_2(l/a)
        +O(1),
\end{equation}
where $c$ is as above the central charge, $l$ is the length of a 
single interval forming region $I$, $a$ is an {\it ultraviolet cutoff}, 
corresponding to a lattice spacing, to avoid an ultraviolet 
divergence, cp.\ eq.\ (\ref{KGLogNeg}). The above
constant $C$ is hence nothing but $c/3$.
This divergence is also removed by using the mutual 
information \cite{Casini}, see Section \ref{MutualInfo}. The 
offset constant in eq.\ (\ref{confent}) is non-universal.  So 
the logarithmic divergence of the entanglement entropy in the 
length of the interval is recovered here. From the expression given 
in ref.\ \cite{Calabrese} for $\rho_I^n$, one also finds for the 
Renyi entropies for $\alpha>1$
\begin{equation*}
        S_\alpha (\rho_I) = \frac{c}{6}\left(1+1/\alpha \right)\log_2(l/a) +O(1).
\end{equation*}
If one is close to the critical point, where the correlation 
length $\xi>0$ is large but finite, one can often still effectively 
describe the system by a conformal field theory. One then obtains
for the entanglement entropy \cite{Calabrese} 
(compare also ref.\ \cite{HuertaJSTAT})
\begin{equation*}
        S(\rho_I) \rightarrow \frac{c}{3} \log_2(\xi/a).
\end{equation*}

\subsection{Disordered spin chains}

Natural systems will generally exhibit a certain amount of 
{\it quenched disorder} which means that the model parameters 
are drawn randomly and the resulting correlation functions 
or entanglement entropies ${\ee}[S(\rho_I)]$ have to be 
considered as being averaged over the a priori distributions, 
with average $\ee$. 
The critical behavior of quantum spin chains with ``quenched''
disorder is remarkably different from its counterpart in the 
corresponding pure case, in several respects. Hence, it is 
only natural to ask whether the scaling of the entanglement 
entropy is influenced by having some disorder in the model. 
This question has first been posed in ref.\ \cite{Refael}
for the spin-$1/2$ {\it random anti-ferromagnetic Heisenberg model},
\begin{equation*}
        H= \sum_{j\in L} J_j\left(
        \sigma^x_j \sigma^x_{j+1}+
        \sigma^y_j \sigma^y_{j+1}+
        \sigma^z_j \sigma^z_{j+1}
        \right),
\end{equation*}
with $\{J_j\}$ drawn from a suitable continuous distribution. The 
low energy properties of this model, along with the random 
XX model, are described by what is called a {\it random-singlet 
phase} \cite{Laflorencie}. 
Using a real-space renormalization group approach 
\cite{Refael}, the intuition can be developed that in this 
phase, singlets form in a random fashion, distributed over 
all length scales. The entanglement entropy of a sub-block 
is hence obtained by effectively counting the singlets that 
cross the boundary of the sub-block.  This intuition is
further developed in ref.\ \cite{Refael2}. Within the framework of
a real-space renormalization group approach---it is shown 
that the averaged entanglement entropy for a large class of 
disordered models scales like
\begin{equation*}
        \ee[S(\rho_I)]= \frac{\gamma}{3}\log_2(n)+O(1).
\end{equation*}
In this class one hence observes universal behavior in the 
scaling of the averaged entanglement entropy. The intuition 
elaborated on above is further corroborated by work on the  
{\it random anti-ferromagnetic XXZ chain} \cite{Hoyos}, 
described by a Hamiltonian
\begin{equation*}
        H=\sum_{j\in L} J_j\left(
        \sigma^x_j \sigma^x_{j+1} 
        +
        \sigma^y_j \sigma^y_{j+1}
        + \Delta_j
        \sigma^z_j \sigma^z_{j+1}
        \right),
\end{equation*}
where $\{J_j\}$ are positive, uncorrelated random variables 
drawn from some probability distribution and the uncorrelated 
anisotropy parameters $\{\Delta_j\}$ are also taken from a 
probability distribution. 
In this work, the observation is further explored that the
scaling of the averaged entanglement entropy can be universal, 
even if correlation amplitudes are not, in that they would 
manifest themselves only in non-leading order terms in the 
entanglement entropy.  This intuition is also further corroborated
in refs.\ \cite{Igloi,HaasDisorder}, on the entanglement entropy in a
2-D situation.

From a fully rigorous perspective, the entanglement entropy 
in the random Ising model---for which ref.\ \cite{Refael} finds a scaling
with the effective central charge of $\ln(\sqrt{2})$--- has recently been revisited with 
methods and ideas of {\it percolation theory} \cite{Grimmett}. 
This approach is more limited than the Fisher-Hartwig techniques
in terms of the class of models that can be 
considered---the Ising model only---but is more powerful in 
that also disordered systems with no translational invariance 
can be considered. 

\begin{theorem}[Non-translationally invariant Ising model]
Consider the
Ising model
\begin{equation*}
        H=- \frac{1}{2}\sum_{j,k\in L,\text{dist}(j,k)=1}
        \lambda_{j,k}X_j  X_k - \sum_{j\in L} \delta_j Z_j,
\end{equation*}
where $\lambda_{j,k}\geq 0$ and $\delta_j\geq0$ are the spin-coupling and 
external field
intensities, respectively, which may depend on the 
lattice site in a non-translationally
invariant system. The total number of sites is
$N=2m+n+1$, with $\{1,\dots, n\}$ being the distinguished
region. Then there exist $\gamma,\alpha,C$ with
properties as in the subsequent footnote 
 \footnote{Let $\lambda,\delta\in(0,\infty)$ and write $\theta = \lambda/\delta$.
There exist constants $\alpha,C \in(0,\infty)$---depending on $\theta$ only---
and a constant $\gamma=\gamma(\theta)$ satisfying
$0<\gamma<\infty$ if $\theta<1$, such that, for all $n\ge 1$,
\begin{equation}\label{eq:rcbound}
        \|\rho_N^n-\rho_{M}^n\| \le \min\{2, C n^\alpha 
        e^{-\gamma N}\},
        \qquad 2\le N\le M.
\end{equation}
Here $\rho_N^n$ denotes the reduced state of $n$ sites in a system of
total size $N$. One may find such a $\gamma$ satisfying $\gamma
\rightarrow \infty$ as 
$\theta\searrow 0$.}.
If $\gamma>4 \ln 2$, then there exist constants
$c_1,c_2>0$ depending only on $\gamma$ such that
\begin{equation}
        S(\rho_I) \leq c_1 \log_2 (n)+c_2, 
\end{equation}
for $m\geq 0$, so the entanglement entropy 
is at most logarithmically divergent.
\end{theorem}

The general picture that emerges is that the entanglement 
entropy scales as in the non-random case, but with a different
prefactor in the logarithmic divergence. This seems natural, 
as the disorder tends to ``localize'' excitations, and hence, 
with faster decaying correlations one would expect less 
entanglement to be present in the system. Yet, there are 
exceptions: Cases in which one does find a logarithmic 
divergence, but with a larger prefactor compared to the 
non-random case. This includes the random quantum 
Potts model with spin dimension $d$: Here, 
for the very large dimension 
of $d>41$ one finds a larger factor \cite{Potts}. 
The exploration and complete classification of the role 
of disorder to entanglement properties of ground 
states---including non-critical and higher-dimensional 
models---remains an interesting challenge.

\subsection{Matrix-product states}

Matrix-product states (MPS) play a very central role in the 
context of area laws for the entanglement entropy. They form 
the class of states that is at the root of the workhorse of 
simulating strongly correlated quantum chains --- DMRG. This 
link will be elaborated upon in detail in Section \ref{Simulate}. Here,
we focus on the entanglement and correlation properties of 
MPS. In the original sense, MPS are states defined on quantum 
chains consisting of $N$ sites, each constituent being a 
$d$-level system. There are several ways of defining and 
introducing MPS, the relationship of which may not be entirely 
obvious. This is also the reason that it was left unnoticed
for some time that MPS--- as being generated in DMRG---and 
{\it finitely correlated states} \cite{FCS}---as being 
considered in the mathematical physics literature---are up 
to translational invariance essentially the same objects.

One way of looking at MPS is via a {\it valence-bond picture}: 
For each of the constituents one introduces a virtual 
substructure consisting of two particles. Per site with Hilbert space $\cc^d$, one associates 
a Hilbert space $\cc^D\otimes\cc^D$ for some $D$. This $D$ is 
sometimes referred to as the dimension of the {\it correlation space}, 
or $D$ called the {\it auxiliary or virtual dimension}. 

\begin{figure}[hbt]
\includegraphics[width=.8\columnwidth]{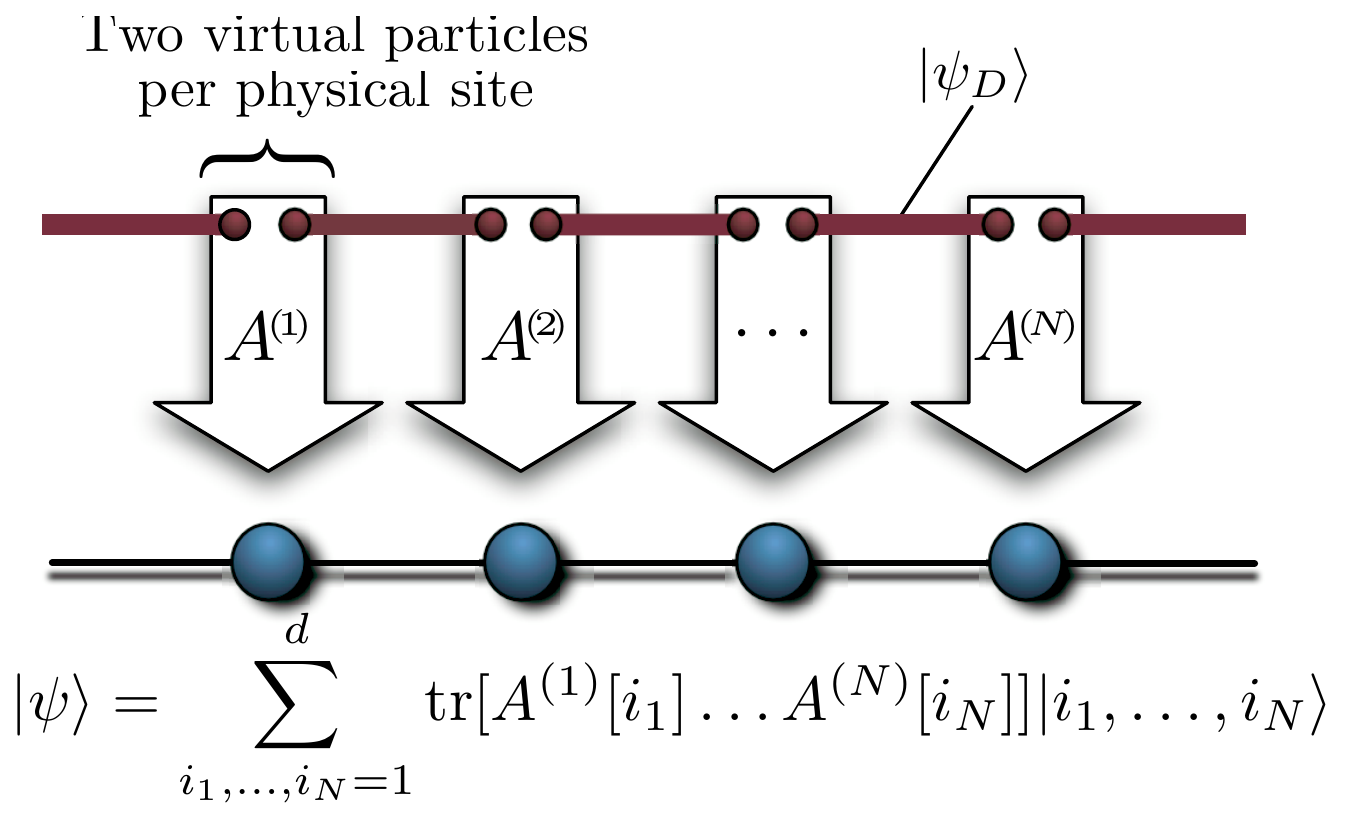}
\caption{The valence-bond picture underlying matrix-product 
states as explained in the main text.\label{VBMPS}}
\end{figure}

These $D$-dimensional virtual systems are thought to be prepared to 
a maximally entangled state with each one particle of each of the 
neighbors, arranged on a ring (see Fig.\ \ref{VBMPS}). In other words, 
one starts from a 
pure state defined by the state vector $|\psi_D\rangle ^{\otimes n}$, 
where we have defined the maximally entangled state vector as 
\begin{equation}\label{maxent}
        |\psi_D\rangle= \frac{1}{\sqrt{D}}\sum_{k=1}^D |k,k\rangle.
\end{equation}  
Then, one applies a local linear map to each of the pairs of 
systems associated with every physical constituent in the center 
of the chain,
\begin{equation}\label{project}
        A^{(k)}= \sum_{j=1}^d \sum_{a,b=1}^D A^{(k)}_{a,b}[j]
        |j\rangle\langle a,b|,
\end{equation}
where $k=1,\dots, N$. This procedure will prepare a certain 
class of states: Indeed the MPS. We may conceive
for each site $k\in L$ the collection of complex
numbers $A^{(k)}_{a,b}[j]$ as the elements of $d$ matrices
$A^{(k)}[1],\dots, A^{(k)}[d]$. For a quantum spin chain with
$d=2$, we hence simply  have two matrices $A^{(k)}[1],A^{(k)}[2]$ per site.
This procedure of locally projecting to the physical dimension $d$
gives rise to state vectors of the form
\begin{equation}\label{MPS}
        |\psi\rangle = \sum_{i_1,\dots, i_N=1}^d
        \text{tr}[A^{(1)}[i_1]\dots A^{(N)}[i_N] ]
        |i_1,\dots, i_N\rangle.
\end{equation}
This is the most frequently used form of representing
{\it matrix product states}. For open boundary conditions 
in a chain $L= \{1,\dots, N\}$, $ A^{(1)}[i_1]$ and $A^{(N)}[i_N] $ 
are row and column vectors, respectively. MPS are described 
by only a number of parameters polynomial in the system size, 
$O(d ND^2)$, in contrast to the scaling of the dimension of 
the full Hilbert space $(\cc^{d})^{\otimes N}$, which is 
exponential in $N$.\footnote{If one allows $D$ to 
(exponentially) grow with the system size, one can easily show 
that actually every state vector from $(\cc^{d})^{\otimes N}$ 
can be represented as a MPS of the form as in eq.\ (\ref{MPS}). 
It is important to note that MPS can not only be described with 
linearly many parameters in the system size: One can also 
efficiently compute local properties from them, which is a property 
not merely following from the small number of parameters 
to define them. For expectation values of
 observables having a non-trivial support on sites $k,\dots, k+l\in L$,
 we find
 $ \langle S_k \dots S_{k+l}\rangle = \text{tr}[E_\id^{(1)}\dots E_\id^{(k-1)}
         E_{S_{k}}^{(k)}\dots E_{S_{k+l}}^{(k+l)} E_\id^{(k+l+1)}\dots E_\id^{(N)}]$,
 where the {\it transfer operators} are defined as
 \begin{equation*}
         E_S^{(l)}= \sum_{j,k=1}^d \langle j|S|k\rangle 
         \left( A^{(l)}[k]\otimes  (A^{(l)}[j])^\ast \right),
 \end{equation*}
 the star denoting complex conjugation.
The decay of correlation functions can also be studied:
If all matrices are the same per site, $A^{(k)}[j]=:A[j]$
 for all $j\in L$, and similarly define
 $E_\id$, then one finds
 \begin{equation*}
 	\langle S_k S_{k+l}\rangle - \langle S_k \rangle  \langle S_{k+l}\rangle  =
         O(  |\lambda_2(E_\id) |^{l-1}),
  \end{equation*}        
where $\lambda_2(E_\id)$ denotes the second to largest eigenvalue of
 the transfer operator of the identity $E_\id$.
}

The particular construction of MPS immediately shows
that MPS satisfy an area law. In fact, it follows trivially 
from their definition (see also fig. \ref{VBMPS}) that
\begin{equation*}
        S(\rho_I) \leq 2 \log_2 (D), 
\end{equation*}
so the entanglement entropy is always bounded from above by 
a constant in $N$. MPS have hence an in-built area law property. 
As we have shown earlier the ground states of a variety of 
Hamiltonians exhibit exactly such an area-scaling when the 
system is non-critical but a logarithmic divergence when the
system is critical. This already suggests that MPS may be
a good description for ground states of non-critical systems but
that this description may become less efficient in critical systems.
Indeed, it will be discussed and highlighted later in this
article that systems satisfying an area law can be economically 
represented as MPS so that MPS with a small auxiliary 
dimension $D$ can indeed typically approximate 
ground states of local Hamiltonians. 

\subsection{Single-copy entanglement}
The entanglement entropy---occupying center stage in this 
article---quantifies entanglement in a very  precise sense: 
For pure states it is the {\it distillable entanglement} \cite{Horodecki,InHouseReview},
so the rate with which 
one can locally extract maximally entangled pairs from a 
supply of identically prepared system. Specifically, local 
refers here to a subsystem $I$ of the system, but to a collective 
operation on many identically prepared states. In a quantum 
many-body system, needless to say, this means that one performs 
operations that are local to all constituents in $I$ collectively 
in all specimens at hand. 

When having the entanglement content in mind, one can equally 
reasonably ask how much entanglement is contained in a {\it single} 
spin chain. The concept of single-copy entanglement 
grasps this notion of distilling entanglement from a single 
specimen of a quantum spin chain with certainty.

If  $D$ is the largest integer such that one can deterministically transform a state into the maximally entangled
state $|\psi_D\rangle\langle \psi_D|$ (see eq.\ (\ref{maxent})) by local operations and classical communication (LOCC), 
i.e.
\begin{equation}\label{SC}
        \rho\mapsto |\psi_D\rangle\langle \psi_D|,
\end{equation}
one assigns the value $E_1=\log_2(D)$ to the state as its {\it single-copy entanglement}.
For pure states, such transformations on the level of specimens 
are perfectly well understood \cite{Nielsen,VidalMaj,JonathanPlenio}
and is linked to the well-established theory of majorization in 
linear algebra \cite{HornJohnson}. For our present purposes, for 
a pure state $\rho=|\psi\rangle\langle\psi|$, we find that
eq.\ (\ref{SC}) holds if and only if 
        $\|\rho_I \|\leq 1/D$.
Hence,
\begin{equation*}
        E_1(\rho_I)= \log_2(\lfloor \| \rho_I\|^{-1}\rfloor).
\end{equation*}
This, in turn, means that the single-copy entanglement can be derived
from the $\alpha$-Renyi entropy of the reduction in the limit of large $\alpha$.
A quite surprising insight is that in critical systems, we do not only find a local
spectrum leading to the logarithmic divergence of the entanglement entropy.
But that there is more structure to the spectrum, governing all of its Renyi
entropies. For example, for quasi-free models, we find that once the entanglement
entropy diverges, so does the single-copy entanglement, with a prefactor
that is asymptotically exactly half the value of the entanglement entropy \cite{Single,Orus}.

\begin{theorem}[Single-copy entanglement]
\label{SingleCopy}
Consider a family of quasi-free fermionic
Hamiltonians as in Theorem
\ref{CriticalLogarithmic}. Then, whenever
the entanglement entropy scales as
\begin{equation*}
        S(\rho_I) = \xi \log_2(n) + O(1),
\end{equation*}
for some constant $\xi>0$,
then the scaling of the single-copy entanglement
is found to be
\begin{equation*}
        E_1(\rho_I)= \frac{\xi}{2} \log_2(n) + O(1).
\end{equation*}
\end{theorem}

This means that exactly half the entanglement can be
distilled from a single critical chain than what is available 
as a rate in the asymptotic setting \cite{Orus,PeschelSingle}).
This finding has also been corroborated by the behavior of all 
critical models for which the local spectra can be described 
by their conformal field theory in quite some generality \cite{Orus}.
Ref.\ \cite{BarthelBoundary}
studies Renyi entropies in 
{\it boundary critical phenomena}, and hence also  arrives
at a relationship between the entanglement entropy and 
the single-copy entanglement.  Ref.\ \cite{LatorreReview}
considers the
entropy loss along the renormalization group trajectory 
driven by the mass term in free massive theories, and 
discusses also the single-copy entanglement in such 
situations. Ref.\ \cite{Keyl}
studies the situation of single-copy entanglement in the
situation of bipartite systems between blocks when there is a 
gap of a finite number of sites between the two blocks. Interestingly,
there are critical models in which the single-copy entanglement still diverges 
in this sense.\footnote{Note that the single copy
entanglement still grasps bipartite entanglement in a
quantum many-body system. Identifying
scaling laws for genuine multi-particle entanglement 
is an interesting enterprise in its own right. Notably,
refs.\ \cite{BoteroMulti,OrusMulti} consider the
{\it geometric entanglement} (the logarithm of the
largest Hilbert-Schmidt scalar product 
with a pure product state) and relate it to the conformal
charge of the underlying model.}
 
\subsection{Summary of one-dimensional systems}

In a nutshell, the situation in one-dimensional translation
invariant models is quite clear: If a system is local and 
gapped, an area law always holds rigorously. In many specific 
models, prefactors can be computed. In contrast, if the 
interactions may be long-ranged
area laws may be violated. For critical lattice models for 
which one can directly evaluate the entanglement entropy, a 
logarithmic divergence is encountered. This picture is 
supported by the findings of conformal field theory. 
The situation will be less transparent and more intricate 
in higher-dimensional models. In any case, in the light of 
the previous findings, one may be tempted to formulate the 
following conjecture on the numerical bound on the right hand
side of the previously discussed area law:

\begin{conjecture}[Area bound in one-dimension] There exists a 
function $f:\rr^+\rightarrow \rr^+$---equipped with further suitable
properties---such that 
in any gapped one-dimensional model, we have  
\begin{equation*}
        S(\rho_I) \leq f({v}/{\Delta E}),
\end{equation*}
where $\Delta E$ is the spectral gap and $v$ is the speed 
of sound as used in the Lieb-Robinson bound.
\end{conjecture}

Indeed, while most explicit studies do indicate a behavior 
linear in $\ln(1/{\Delta E})$
of the entanglement entropy---the above mentioned quasi-free models---one 
can construct models 
\cite{GottesmanHastings,Irani}
for which 
one finds a dependence 
which is polynomial in $1/{\Delta E}$.

\section{Area laws in higher dimensions}

For a chain, to satisfy an ``area law'' for the 
entanglement entropy means simply that it saturates with 
increasing block sizes. Needless to say, the notion of 
having entropic quantities scaling like the boundary area 
of a subregion becomes specifically relevant in case of 
higher dimensions: Then the boundary of the region $I$ is 
a non-trivial object. Now we are in a position to 
approach the question: Given a ground state of a quantum
many-body system, does the entanglement entropy of a subregion 
$I$ fulfill an area law? This question has been initiated 
in refs.\ \cite{Bombelli,Srednicki}, where also a numerical 
answer has been found.

The answer to this question for ground states is very much 
developed in case of---once again---quasi-free bosonic or 
fermionic models. Even in such systems, the rigorous answer 
to this question will turn out to be technically quite involved. 
The reason for these technicalities is essentially rooted in 
the very fact that one distinguishes a subregion, thus, e.g., breaking
translational symmetry of translationally invariant systems, and analytical methods 
are hard to come by. The first rigorous higher-dimensional 
area law has been proven---to the knowledge of the 
authors---in refs.\ \cite{Area}, with refinements for 
arbitrary harmonic interactions in ref.\ \cite{Area2,Area3}
so that for such bosonic free models the problem can be
considered solved in all generality forming a ``laboratory'' 
of what one should expect in general systems. 

For critical fermionic models \cite{Wolf,Klich,Halfspace,Fermi2d,FarkasLong,MustCite},
quite intriguingly, one can find small violations of area laws: 
The area law is then only satisfied up to a logarithmic correction.
In this section, we will discuss quasi-free models in great detail.
Beyond such quasi-free models, no rigorous results are known for
states at zero temperature, with the exception of classes of states that satisfy an 
area law by their very construction, and a subsection will specifically be devoted 
to those. 

The models discussed here, however, do provide a clear intuition:
Whenever one has a gapped and local model, and hence a length scale
provided by the correlation length, one should reasonably expect 
an area law to hold. 
In cases where the number of eigenstates with 
vanishing energy-density is not exponential in the volume - a
technical condition in its own right - one can even prove an area 
law with at most a logarithmic correction from a sufficient 
decay of correlations \cite{MasanesArea}.
The converse is not true, as we will see later, 
and one can have area laws even for critical systems in which the 
correlation length does not provide a length scale. For systems 
at non-zero temperatures, by contrast, the entropy of entanglement 
neither forms a meaningful measure of entanglement nor for quantum 
correlations. For appropriately defined measures of correlations,
however, one can restore an area law which holds in generality for 
a large class of systems.

\subsection{Quasi-free bosonic and fermionic models: 
Sufficient conditions for an area law}

We will follow the general description of ref.\ \cite{Area3}, 
where we think of the model being defined on a general lattice 
$L$ specified by a general simple graph. We consider quadratic 
bosonic Hamiltonians as in eq.\ (\ref{BosonHamilton}) and quadratic fermionic 
Hamiltonians as in eq.\ (\ref{GeneralFermionic}).
The key step is to relate correlation functions to entropic 
quantities. As before in the case of a harmonic chain, it is 
very involved to think of the entropy of entanglement itself. 
What comes to our help, however, is again that we can use
the logarithmic negativity as an upper bound to the entanglement 
entropy (see eq.\ (\ref{logengineq})). The logarithmic negativity
is easier to treat analytically, as we can at all times refer 
to the full system, and not to subsystems $I$. In fact, we 
find that the logarithmic negativity can be bounded from above 
by the $L_1$-norm of a submatrix of the covariance matrix \cite{Area}.
For fermions in turn, the entropy may be bounded 
directly using the bound in eq.\ (\ref{fbound}).

\begin{theorem}[Entropic bounds from matrix norms]
The entanglement entropy of ground states of quadratic bosonic 
Hamiltonians as in eq.\ (\ref{BosonHamilton}) satisfies 
\begin{equation*}
        S(\rho_I)\leq E_{N}(\rho,I)\leq 8\|\Gamma_x\|
        \sum_{i\in I, j\in O}\left|\langle p_ip_j\rangle\right| .
\end{equation*}
The entanglement entropy of unique ground states of quadratic 
fermionic Hamiltonians as in eq.\ (\ref{GeneralFermionic}) 
satisfies 
\begin{equation*}
        S(\rho_I)\leq 2\sum_{i\in I, j\in O} 
        |\langle f_i^\dagger f_j\rangle+\langle f_i f_j\rangle |.
\end{equation*}

\end{theorem}

This is a key tool towards proving the main theorem: We can 
reduce the evaluation of an entropic quantity to a counting 
argument over terms that can be evaluated from two-point 
correlators. Note that the use of the logarithmic negativity 
results in an important simplification of the problem. This 
shows that ideas from quantum information theory indeed help in 
finding proofs of statements of the scaling of the entanglement 
entropy. 

We are now in the position to state the bound of the scaling of 
the ground state entanglement in the boundary area $s(I)$, 
eq.\ (\ref{BoundaryArea}),  of the distinguished region 
$I$ \cite{Area2,Area3,marcusThesis}.
It is remarkable that merely the decay of two-point correlations
matter here, and that even some critical models will give rise to an
area law, as long as the algebraic decay of correlations is sufficiently strong.
\begin{theorem}[Quadratic Hamiltonians on general lattices]
Let $\eta= \mathcal{D}+1+2\varepsilon$, $\varepsilon >0$, and 
assume that the ground state is unique and fulfills for 
$i,j\in L$, $i\ne j$, and some constant $K_0$
\begin{equation*}
\frac{K_0}{dist^\eta(i,j)} \ge\begin{cases}
|\langle p_ip_j\rangle| &\text{ for bosons,}\\
|\langle f_i^\dagger f_j\rangle+\langle f_if_j\rangle | &\text{ for fermions.}
\end{cases}
\end{equation*}
Then
\begin{equation*}
        S(\rho_I)\le K_0c_{\mathcal{D}}\zeta(1+\varepsilon)
        s(I)\times\begin{cases}
        \|\Gamma_x\|&\text{ for bosons},\\
1 &\text{ for fermions},
\end{cases}
\end{equation*}
where $\zeta$ is the Riemann zeta function and the 
constant $c_{\mathcal{D}}$ depends only on the dimension 
of the lattice. 
\end{theorem}

A general version of what one should expect to be true provides 
the connection to the spectral gap: For gapped models the correlation
functions decay exponentially with the graph theoretical distance. 
One cannot apply the 
Lieb-Robinson Theorem to prove this, unfortunately, as the 
involved operators are unbounded. Hence, a technique that is 
applicable to describe clustering of correlations in such models 
had to be developed. The ideas of the proof go back to ref.\ 
\cite{Benzi}, generalized to arbitrary lattices 
in ref.\ \cite{Area3,marcusThesis}. Key ideas of the 
proof are polynomial approximations in the sense of Bernstein's 
Theorem. For a thorough discussion of clustering of correlations 
in translation-invariant harmonic systems, see ref.\ \cite{Schuch}. 
For general lattices and gapped quadratic bosonic and fermionic 
Hamiltonians, one finds that two-point correlation functions 
decay exponentially. Together with the above theorem this leads 
to an area law whenever the model is gapped:

\begin{corollary}[Area law for gapped quasi-free models]
The entanglement entropy of ground states of 
local gapped models of the type of
eq.\ (\ref{BosonHamilton}) for bosons and of
eq.\ (\ref{GeneralFermionic}) for fermions for arbitrary lattices $G=(L,E)$ and
arbitrary regions $I$ 
satisfies for a suitable constant $\xi>0$
\begin{equation*}
        S(\rho_I)\leq \xi s(I).
\end{equation*}

\end{corollary}

\subsection{Logarithmic correction to an area law: Critical fermions}

What can we say about situations in which the previous
sufficient conditions are not satisfied? Specifically, how is the 
scaling of the entanglement entropy modified in case of 
critical fermionic models? This is the question that will feature
in this subsection. Following the bosonic result in refs.\ \cite{Area,Area3},
the entanglement entropy in fermionic models was first studied in
ref.\ \cite{Wolf} for cubic lattices. 
Here, the quadratic bound in eq.\ (\ref{fbound})
plays an important role, to relate bounds to the entropy to feasible
expressions of the covariance matrix of the ground state. 
Here, not quite an area law, but only one up to a logarithmic
correction is found. The results can be summarized as follows:

\begin{theorem}[Violation of area laws for critical fermions]
For a cubic sublattice $I=\{1,\dots, n\}^{\times {\cal D}}$
and an isotropic quasi-free model as in eq.\
(\ref{GeneralFermionic})
with a Fermi sea of
non-zero measure and a finite non-zero surface
there exist constants $c_0,c_1>0$ such that
the ground state fulfils
\begin{equation*}
       c_0  n^{{\cal D}-1}\ln(n)\leq 
       S(\rho_I)\leq
       c_1 n^{{\cal D}-1}\ln^2(n).
\end{equation*}
\end{theorem}
The stated lower bound makes use of 
the assumption that the Fermi surface is finite
(and of a technical assumption that the 
sets representing the states 
cannot have nontrivial irrelevant  directions);
assumptions both of which can be 
removed \cite{FarkasLong}.

This fermionic quasi-free case already exhibits a quite 
complex phase diagram \cite{Fermi2d,MustCite}.
At the same time, ref.\ \cite{Klich} formulated a similar 
result, based on a conjecture on the validity of 
Fisher-Hartig-type scaling result for higher dimensional equivalents
of Toeplitz matrices, as was further numerically corroborated in ref.\ \cite{MustCite}. 
A  logarithmic divergence is not directly inconsistent
with the picture suggested in a conformal field theory setting, as 
relativistic conformal field theories do not have a Fermi surface \cite{Japanese}.
It is still intriguing that critical fermions do not
satisfy an area law, but have logarithmic corrections. In this
sense, critical fermionic models could be said to 
be ``more strongly entangled'' than critical bosonic 
models.

\subsection{Difference between critical fermions and bosons: 
Half spaces}

The scaling of block entropies for bosons and fermions in higher 
spatial dimensions hence 
exhibit remarkable differences. Let us consider 
the case of a cubic lattice of $n^{\cal D}$ sites with periodic boundary 
conditions and $I=\{1,\dots, m\}\times\{1,\dots,n\}^{\times\mathcal{D}-1}$ 
(w.l.o.g. we distinguish the first spatial dimension). Then one may 
transform the Hamiltonian to a system of mutually uncoupled 
one-dimensional chains using a unitary discrete Fourier transform. 
After this decoupling procedure the entanglement between $I$ and 
$O$ is given by a sum of the entanglement between the sites 
$I^\prime=\{1,\dots, m\}$ and $O^\prime=\{m+1,\dots,n\}$ of 
the $n^{\mathcal{D}-1}$ individual chains 
\begin{equation*}
        \frac{S(\rho_I)}{s(I)} = \frac{1}{n^{\mathcal{D}-1}}
        \sum_{i=1}^{n^{\mathcal{D}-1}}S(\rho^i_{I^\prime}).
\end{equation*}
We start with a discussion of fermions and focus on the isotropic 
setting ($B=0$ in eq.\ (\ref{GeneralFermionic})). After taking the 
limit $n\rightarrow\infty$, the asymptotic behavior in $m$ of the 
entanglement $S(\rho^i_{I^\prime})$ can then be read off Theorem 
\ref{CriticalLogarithmic} to yield the following statement (for technical details see ref.\ 
\cite{Halfspace}):

\begin{theorem}[Prefactor for fermionic half spaces] Asymptotically,
the entanglement entropy of fermionic isotropic models of half spaces
satisfies
\begin{equation*}
\lim_{n\rightarrow\infty}\frac{S(\rho_I)}{s(I)}=
\frac{\log_2 (m)}{6}\sum_{s=1}^\infty sv_s+O(1).
\end{equation*}
Here,
\begin{equation*}
v_s=\frac{\left|\left\{\phi\in [0,2\pi)^{\mathcal{D}-1}\,
\big|\, \sigma(\phi)=s\right\}\right|}{(2\pi)^{D-1}}
\end{equation*}
is the integral over individual chains $\phi$ with $s$ discontinuities 
$\sigma(\phi)$ in their symbol.
\end{theorem}

 Hence, one encounters a logarithmic divergence 
in $m$ of the entanglement entropy and the pre-factor depends on 
the topology of the Fermi-surface: The symbols exhibit discontinuities 
on the Fermi-surface. If the Fermi surface is of measure zero (i.e.,
the set of solutions to $\lambda_{\phi}=0$, $\phi\in[0,2\pi)^{\mathcal{D}}$, 
is countable, as, e.g., in the critical bosonic case discussed below), 
we have $v_s=0$ and the system obeys the area law. 

For the bosons, we discuss the important case of $m=n/2$ for the 
Klein-Gordon Hamiltonian as in eq.\ (\ref{KG}).  After the transformation
to uncoupled chains, one finds Hamiltonians for the individual chains 
that correspond to a nearest-neighbor coupling matrix $X$ of the form 
as in Theorem \ref{T1}, which yields
\begin{equation*}
\frac{E_{N}(\rho,I)}{s(I)}=\int\!\!\frac{\md\varphi}{4(2\pi)^{D-1}}\log_2\left(\frac{\mathcal{D}-\sum_{d=1}^{\mathcal{D}-1}\cos(\varphi_d)+1}{\mathcal{D}-\sum_{d=1}^{\mathcal{D}-1}\cos(\varphi_d)-1}\right)
\end{equation*}
in the limit $n\rightarrow\infty$. This expression is independent of 
the mass $m$ and finite: For $\mathcal{D}=2$, it evaluates to 
$\log_2(3+2\sqrt{2})/4$ and similarly for $\mathcal{D}>2$. Hence, 
despite being critical, the system obeys an area law, in contrast 
to the fermionic case (for $m=n/2$ the entanglement for a critical 
fermionic system would diverge in $n$).

Hence, in quasi-free critical models, it matters whether a system is
bosonic or fermionic when it comes to the question whether or not
an area law holds. The above results confirm the numerical analysis
of ref.\ \cite{Srednicki} for critical bosonic theories, and of
ref.\ \cite{Fermi2d} for two-dimensional fermionic systems. Motivated 
by these findings, ref.\ \cite{Subarea}  numerically 
studies the non-leading order terms of an area law in nodal fermionic 
systems: It is found that in non-critical regimes, the leading subarea 
term is a negative constant, whereas in critical models one encounters
a logarithmic additive term. A lesson from these higher-dimensional
considerations is that the simple relationship between criticality 
and a violation of an area law is hence no longer valid for local 
lattice models in $\mathcal{D}>1$.

\subsection{Entanglement in bosonic thermal states}\label{AreaThermal}

In this subsection, we briefly discuss area laws for notions of 
entanglement in Gibbs states,
\begin{equation*}
        \rho_\beta=\frac{\exp(-\beta {H})}{\text{tr}[\exp(-\beta {H})]}
\end{equation*}
for some inverse temperature $\beta>0$. The second moments matrix, the 
covariance matrix, is then found to be $\Gamma=\Gamma_x\oplus\Gamma_p$ 
\cite{Area2}
\begin{equation*}
\begin{split}
\Gamma_x&=X^{-1/2}\bigl(X^{1/2}PX^{1/2}\bigr)^{1/2}(\id + G)X^{-1/2},\\
\Gamma_p&=X^{1/2}\bigl(X^{1/2}PX^{1/2}\bigr)^{-1/2}(\id + G)X^{1/2},\\
G&=2\Bigl(\exp\bigl[\beta\bigl(X^{1/2}PX^{1/2}\bigr)^{1/2}\bigr]-\id\Bigr)^{-1}.
\end{split}
\end{equation*}
Using the the methods of refs.\ \cite{Benzi,Area2} one again finds 
the suitable decay of correlations, which can be translated into 
an area law for the entanglement content. Here, the result---taken 
from refs.\ \cite{Area2,Area3}---is stated in terms of the logarithmic 
negativity. 

\begin{theorem}[Entanglement in thermal bosonic states]
The logarithmic 
negativity of thermal states of quadratic finite-ranged
bosonic Hamiltonians as in eq.\ (\ref{BosonHamilton}) for $[X,P]=0$ satisfies
$E_{N}(\rho,I)\leq \xi s(I)$ for a suitable constant $\xi>0$.
\end{theorem}

Since the logarithmic negativity is an upper bound to the
{\it entanglement of formation} and hence the
{\it distillable entanglement} \cite{Horodecki,InHouseReview}
this implies an area law for these quantities as well.
It is important to stress that the entropy of a subregion 
as such no longer reasonably quantifies entanglement between
that subregion and the rest of the lattice: Even classically 
correlated separable states will in general have a positive 
entropy of the reduced state. The latter quantity is then 
indeed extensive and fulfills a volume law, unlike the entanglement 
content. Area laws in thermal states have further been studied 
in detail in ref.\ \cite{Janet}, where an emphasis has been put
on identifying regions where the states become separable.
Refs.\ \cite{AcinThermal,AcinThermal2} 
investigate thermal bound entanglement---entanglement that is not
distillable---in bosonic quadratic and spin systems. 

\subsection{Results from conformal field theory}

In systems with more than one spatial dimension, the situation is
more intricate, and there is no general expression known
for entanglement entropies in $d+1$-dimensional conformal field
theories. For interesting steps into a description of systems
with two spatial dimensions in the framework of conformal field
theory see refs.\ \cite{Fradkin,Japanese}. 
For a class of critical models in two spatial dimensions 
(including the quantum dimer model), it is found that 
        $ S(\rho_I) = 2 f_s (L/a) + c g \ln(L/a)+O(1)$,
where $L$ is the length of the boundary area, 
$f_s$ is an area law prefactor that is interpreted as a {\it boundary
free energy}, and $g$ is a coefficient that depends on the 
geometric properties of the partition into $I$ and $O$.
That is, in addition to a non-universal area law, 
one finds a universal logarithmically divergent correction. 
For a further discussion of steps towards a full theory of entanglement
entropies in $d+1$-dimensional conformal field theories,
see refs.\ \cite{Fradkin,Japanese}. 

\subsection{States satisfying area laws by construction: Projected
entangled pair states, graph states, and entanglement 
renormalization}\label{PEPS}

In this section, we will discuss classes of states that have the 
area law already built into their very construction. In this sense, 
they grasp the entanglement structure of local higher-dimensional 
models. These are {\it projected entangled pair states}, so 
matrix-product states in higher dimensions, and states from {\it 
entanglement renormalization}. They are designed to be variational 
states well-approximating true ground states of local many-body 
systems: As was already true for matrix product states, they form 
a complete set of variational states. Yet, typically, for a much 
smaller, polynomial or constant, number of variational parameters, 
they often deliver a very good approximation. In projected entangled 
pair states, locality is respected in just the same way as for MPS. 
Entanglement renormalization, in turn, is based on a scale-invariant 
tree structure, intercepted by disentangling steps, which in higher 
dimensions nevertheless leads to an area law for the entanglement 
entropy.

Projected entangled pair states (PEPS) can be thought of as being 
prepared as MPS in higher dimensional cubic lattices 
$L=\{1,\dots, N\}^{\times {\cal D}}$, or in fact to any lattice 
defined by any undirected simple graph $G=(L,E)$. In this valence 
bond construction, one again associates a physical space with 
Hilbert space $\cc^d$ with each of the vertices $L$ of $G$.
Then, one places a maximally entangled pair of dimension $D\times D$ 
(see eq.\ (\ref{maxent})) for some positive integer $D$ between 
any two vertices that are connected by an edge $e\in E$. For a 
cubic lattice, one hence starts from a cubic grid of maximally 
entangled state vectors. 
Then, one applies a linear map $P^{(k)}:{\cc^D}\otimes \dots\otimes 
{\cc^D}\rightarrow \cc^d$ to each physical site, as
\begin{equation*}
        P^{(k)}= \sum_{j=1}^d \sum_{i_1,\dots, i_{|S_1(k)|}}^D
        A_{j,i_1,\dots, i_{e_k}}^{(k)} 
        |j\rangle\langle i_1,\dots, i_{|S_1(k)|}|.
\end{equation*}
Here, $|S_1(k)|$ is the vertex degree of the vertex $k\in L$.
The resulting state vector as such hence becomes 
\begin{equation*}
        |\psi\rangle=  \sum_{i_1,\i_2,\dots, i_{|L|}=1}^d
        {\cal C}[\{A^{(k)}_{i_l} \}_l]
        |i_1,i_2,\dots, i_{|L|}\rangle,
\end{equation*}
where ${\cal C}$ denotes a contraction of all higher-order tensors 
with respect to the edges $E$ of the graph. This amounts to a 
summation over all indices associated with connected vertices. The 
objects $A^{(k)}$ are hence tensors of an order that corresponds 
to the vertex degree of the lattice (a second order tensor---a 
matrix---for a one-dimensional chain, a three order tensor in 
hexagonal lattices, a fourth order tensor in cubic lattices with 
${\cal D}=2$, and so on). This construction is the natural 
equivalent of the valence bond construction for matrix-product 
states as explained in eq.\ (\ref{project}). This ansatz as such 
is the one of {\it tensor product states} that is due to ref.\
\cite{MartinDelgado} which in turn is generalizing earlier work 
on AKLT-type valence bond states in two dimensions in refs.\  
\cite{Hieida,Niggeman}. The generated class of states is referred 
to as {\it projected entangled pair states} \cite{PEPSOld},
reflecting the preparation procedure. PEPS states are sometimes 
also in higher dimensions simply referred to as {\it matrix product 
states} \cite{Try}. This ansatz has proven to provide a powerful 
and rich class of states.  Importantly, ref.\ \cite{PEPSOld} provides
a first simulation method based on PEPS.\footnote{Note that 
1-D MPS based on a suitable order of the constituents
do not form a good approximation for 2-D models. This is essentially
rooted in the observation that one should expect an area law
for the entanglement entropy in gapped 
2-D models. For an important
early discussion of spectra of subsystems in 2-D integrable 
models see ref.\ \cite{Peschel2D} and, for a more recent discussion of 
the implication on DMRG, ref.\ \cite{PEPSOld}.}

This class of states is complete, in that any state of a given 
finite lattice can be arbitrarily well approximated by such a state 
if $D$ is sufficiently large. Clearly, to compute local 
observables in such an ansatz, one has to contract this instance of a 
tensor network which in 2-D is actually computationally 
hard.\footnote{In fact, it is known that the exact
contraction of such a tensor network is contained in the 
complexity class $\#$P-complete \cite{Contraction}. Clearly, 
this means that no algorithm is known with polynomial running 
time.}
It is however possible to provide approximation techniques, 
related to the DMRG approach, that allow for the contraction 
of the tensor network and then for the computation of the 
expectation values of local observables \cite{PEPS,2DHardCoreBosons,2DPEPS,VerstraeteBig}.

A particularly simple yet important subset of the projected entangled 
pair states is constituted by the so-called {\it graph states} 
\cite{Graphs,Schlinge,GraphsLong}: They are instances of {\it stabilizer 
states} \cite{GottesmanThesis,AudenaertP05} which can be thought of 
as being prepared in the following 
fashion: On any graph $G=(L,E)$, one associates each vertex with a 
$\cc^2$-spin. This spin is prepared in $|+\rangle=( |0\rangle+|1\rangle)/2^{1/2}$. 
Then, one applies a {\it phase gate}
\begin{equation*}
        U=|0,0\rangle\langle0,0| + 
        |0,1\rangle\langle0,1|+
        |1,0\rangle\langle1,0|-
        |1,1\rangle\langle1,1| 
\end{equation*}
to each pair of vertices that are connected by an edge. This phase 
gate corresponds to an Ising interaction. Clearly, this construction 
makes sense for any simple graph, and this is a subset of the above 
projected entangled pair states. Graph states readily satisfy an area 
law by construction \cite{Graphs,Zanardi2} as one merely needs to count 
the edges over the boundary of a distinguished region to obtains the 
entanglement entropy, then obviously linear in the 
boundary area.\footnote{The other side of the coin of the difficulty of 
actually contracting tensor networks, even if they correspond to states 
that approximate ground states satisfying area laws well, is that such 
states can have computational power for quantum computing. Indeed, certain 
graph states or {\it cluster states}---as they are called for a cubic 
lattice---are universal resources for quantum computing: Quantum computing 
can be done by merely applying local measurements onto single sites of 
such a cluster states, without the need of additional unitary control. 
This computational model---known as {\it one-way computing} 
\cite{Oneway}---can also be understood as a teleportation scheme in 
virtual qubits \cite{Valence}. The tensor networks that occur when
performing Pauli measurements can still  be efficiently contracted,
but not under arbitrary measurements, leading to universal
computation. The program of using general projected 
entangled pair states in quantum computing based on measurements only 
has been pursued in refs.\ \cite{Models1,Models2}, giving rise to a 
wealth of new {\it measurement-based quantum computational models}. 
This also highlights how the disadvantage of having no classical 
efficient description can be made an advantage: One can at each 
 instance of the computation not efficiently compute the outcome, but 
 on a physical system realizing this model one could efficiently 
 simulate any quantum computer.}

Graph states may be generalized to {\it weighted graph states} 
\cite{Weighted,GraphsLong,PlenioJMO} where the edges may carry a 
different weight, and in turn generalize to the ansatz of a
{\it renormalization algorithm with graph enhancement} (RAGE),\cite{Rage} 
being a strict superset of matrix-product states 
and weighted graph states, one that can nevertheless be efficiently 
contracted. As the graph defining the (weighted)-graph state does
not need to have the same structure as the graph of the physical
system whose quantum state we would like to describe, (weighted)-graph 
states may describe volume scaling on the level of the physical
system. This makes them particularly suitable for simulation of time 
evolution, where no area law can be expected to hold. 

Yet a different class of many-body states with applications in the simulation of 
quantum spin systems is given by the states  generated by {\it entanglement 
renormalization} (MERA) \cite{MERA1}. 
This is a class of states the construction of which is
inspired by a renormalization scheme. Consider a 
tree tensor network with
the physical sites at the end. This can be efficiently contracted. Yet, when decimating, say,
two spins of one layer to a single ``superspin'' in the next layer in a single step of a 
renormalization procedure , one loses information information about the state. The idea
of a MERA ansatz is to allow for {\it disentangling unitaries}, effectively removing
entanglement from a state, before doing a renormalization step.  

More specifically, consider a cubic lattice $L=\{1,\dots, N\}^{\times {\cal D}}$
in some dimension ${\cal D}$, embodying $N^{\cal D}$ sites. Each 
site $j\in L$ is associated with a physical system with Hilbert 
space $\cc^d$. The MERA is essentially a unitary tensor network of depth
$O(\ln(N))$, preparing $|\psi\rangle$ from $|0\rangle^{\otimes N}$. 
It consists of layers of isometries ---performing the renormalization 
step---and disentanglers, which minimize the entanglement in each 
step before the next renormalization step. This renormalization 
step may be labeled with a fictitous time parameter. Each of the 
unitary disentanglers  $U\in U(d^m)$ in the disentangling layer
has a finite support on $m$ sites. In the simplest possible 
realization of a MERA this would be $m=2$. The unitaries can be 
taken to be different in each layer, and also different from each 
other within the layer. Unlike PEPS, they do not give rise to 
strictly translationally invariant states, even if all unitaries 
are taken to be identical in each layer. 
 
Such a procedure can be defined for cubic lattices of any dimension 
${\cal D}$. In ${\cal D}=1$, one does in fact not observe an area law, 
but typically a logarithmic divergence of the entanglement entropy, quite like 
in critical spin systems. Indeed, the MERA ansatz as a scale invariant 
ansatz is expected to be suitable to approximate  critical systems 
well, and numerical simulations based on the MERA ansatz corroborate 
this intuition \cite{Flow,Rizzi,MERA2,EvenblyNew}. 
A precise connection between homogeneous
instances of a MERA ansatz and conformal field theory is established in
refs.\ \cite{MERAC1,MERAC2}. In more than one dimension, 
${\cal D}>1$, MERA satisfy again an area law, as a moment of thought 
reveals: On encounters asymptotically linearly many unitaries over a boundary that 
have entangling power, rendering the computation of an upper bound 
to the entanglement entropy a combinatorial problem. 
Despite this observation, first numerical work on fermionic instances of a
MERA ansatz appear to deliver promising results for strongly correlated
systems \cite{MERAF1,MERAF2,MERAF3,MERAF4}.

\begin{theorem}[Area laws for PEPS, graph states, and MERA]
        For any finite dimension $D$ of the virtual systems, the entanglement entropy 
        of a projected entangled pair state satisfies
        $S(\rho_I)\leq  s(I) D$,
        where as before $s(I)$ denotes the surface area of $I$ on a graph.
        Hence, also graph states with a fixed vertex degree 
        satisfy area laws. A family of states 
        from entanglement renormalization will also satisfies 
        an area law for cubic lattices with ${\cal D}\geq 2$, and a logarithmic
        divergence in ${\cal D}=1$.
\end{theorem}

Interestingly, based on a PEPS description, one can construct critical models
that still satisfy an area law in ${\cal D}=2$ \cite{PEPS},
resembling the situation for critical quasi-free bosonic systems.
The validity of an area law follows trivially from construction, so the
technical part in the argument amounts to showing that a model is
critical. In ref.\ \cite{PEPS} this is shown by employing a 
quantum-classical-correspondence: Take a classical two-body spin Hamiltonian of the
form 
        $H(\sigma_1,\dots, \sigma_N)= \sum_{\text{dist}(i,j)=1}
        h(\sigma_i,\sigma_j)$, 
$\sigma_i =1,\dots, d$. This Hamiltonian will have at
some inverse temperature $\beta>0$ a partition function $Z=\sum_\sigma
e^{-\beta H(\sigma)}$. From this classical partition function, a quantum state
can be constructed  by using the Boltzmann weights as superposition coefficients,
\begin{equation*}
        |\psi_{H,\beta}\rangle=
        \frac{1}{Z^{1/2}} \sum_{\sigma_1,\dots, \sigma_N} e^{-\beta H(\sigma_1,\dots, \sigma_N)/2}
        |\sigma_1,\dots, \sigma_N\rangle.
\end{equation*}
This state vector has the properties that for diagonal observables, it gives rise to the same
expectation values and correlation functions as the corresponding classical model does,
it has a simple representation as a PEPS for $D=d$, and it is---as any PEPS---the
ground state of a local Hamiltonian. The classical model can then be chosen such that
the appropriate decay of correlation functions follows. This construction delivers critical spin
models that nevertheless satisfy an area law.

\subsection{Quenches and non-equilibrium dynamics}\label{noneq}

A physical setting that receives a lot of attention in the recent
literature is the one of non-equilibrium dynamics of quantum many-body systems. 
A specifically interesting setting is the one of a sudden quench
\cite{Dynamical,Calabrese2007,PeschelNew,CalabreseXYQuench,CalabreseExtended,HarmonicDynamical,Relax,Kollath,QuenchDMRG,PeschelQuench,Barthel,CalabreseQuenchEntanglement,CalabresePRL,CoherentMatterWave,Zurek,Quench,Quench3,SchuchQuench,LightCone}: 
Here, the initial condition is the non-degenerate ground state of 
some local Hamiltonian $H$, with state vector $|\psi\rangle$. Then,
one suddenly (locally)
alters the system parameters to a new Hamiltonian $V$.
Since $|\psi\rangle $ will typically no longer be an eigenvalue of 
$H$, one arrives at a non-equilibrium situation: The state vector's 
time evolution is simply given by 
\begin{equation*}
	|\psi(t) \rangle=e^{-\mi tV}|\psi
	\rangle. 
\end{equation*}
Studies of instances of such complex {\it non-equilibrium
many-body dynamics} and questions of the {\it dynamics of quantum 
phase transitions} are enjoying a renaissance recently, not the 
least due to the advent of the high degree of control over quantum 
lattice systems with cold atoms in optical lattices.\footnote{The 
interesting situation
of locally perturbing the {\it state} and hence generating a non-equilibrium
situation has also been considered 
in refs.\ \cite{Calabrese2007,PeschelNew,PeschelQuench}, where 
an area law is always expected to hold.} 

For finite times, infinite {\it quenched systems} satisfy an area 
law in the entanglement entropy 
\cite{CalabreseQuenchEntanglement,Quench,Quench2} (strictly 
speaking whenever one considers time evolution under local 
finite-dimensional Hamiltonians starting from product states).
For finite systems this holds true for times that are sufficiently 
small compared to the system size over the speed of sound.
The intuition is that when suddenly switching to a new Hamiltonian, 
local excitations will be created. These excitations will propagate 
through the lattice, but---except from an exponentially suppressed 
tail---at most with the Lieb-Robinson velocity of 
Theorem \ref{LRT} \cite{Quench2,Decay,LightCone,Quench}.
This is yet again a 
consequence for the approximate locality in quantum lattice systems, 
reminding of the situation in relativity and implies that correlations
can only slowly build up, resulting in an area theorem. In turn, 
such a quench does in general give rise to a linear increase in 
the entanglement entropy, a statement that is provably correct, 
and has been encountered in numerous numerical studies on quenched 
non-equilibrium 
systems \cite{CalabreseQuenchEntanglement,Quench,Quench2,Relax,SchuchQuench,Barthel}.
In fact, finite subsystems can locally relax in time, to appear as 
if they were in a thermal state \cite{Relax}.
These results may be summarized in the following statement.
\begin{theorem}[Area laws in non-equilibrium 
systems]
\label{QuenchTheorem}
Let $|\psi\rangle$ be a product initial state vector, 
and $H$ a local Hamiltonian. Then, for 
any time $t>0$ there exist constants $c_0,c_1>0$ such that for
any subset $I$ the entanglement entropy of the time evolved 
reduction $\rho_I$ of 
$\rho(t)=e^{-\mi tH}|\psi\rangle\langle \psi| e^{\mi t H}$
satisfies
\begin{equation}\label{Increase}
        S(\rho_I(t)) \leq c_0 s(I)  + c_1.
\end{equation} 
Specifically, this is true for any local Hamiltonian on a cubic lattice in dimension
${\cal D}$. This means that for any constant time, 
the entanglement entropy satisfies what is called an {\it area law}. In turn, 
there are product initial state vectors $|\psi\rangle$ of one-dimensional spin chains,
local Hamiltonians $H$, and 
constants $c_2,c_3,c_4, L_0, s_0, t_0 >0$ such that 
\begin{equation*}
        S(\rho_I(t)) \geq c_2 t + c_3,
\end{equation*}
for $L\geq L_0$ and $s\geq s_0$ and $t_0\leq t\leq c_4 s$,
for $I=\{1,\dots, s\}$.
\end{theorem}

That is, for any fixed time $t$, one encounters an area law for the
entanglement entropy, but the prefactor can grow linearly in time. In fact,
by a suitable choice of blocks, one can show that a lower bound
grows linearly in time! This fact is responsible for the hardness
of simulating time evolution of quantum many-body systems using
instances of the DMRG approach: to represent such states
faithfully, exponential resources are then required. Similar bounds give rise
to statements on the minimal time needed in order to prepare
states with topological order using local Hamiltonians \cite{Quench2}.

There is an interesting {\it localization effect} of entanglement under 
quenched disorder, linking to the previous discussion on ground state
entanglement in disordered systems. 
Whereas one obtains from Lieb-Robinson bound the estimate in time
\begin{equation*}
        S(\rho_I(t)) \leq c_0 |t|  + c_1
\end{equation*} 
for suitable constants $c_0,c_1>0$, in the disordered one-dimensional
XY spin chain this bound is replaced by the tighter bound
\begin{equation*}
        S(\rho_I(t)) \leq c_0 \ln(N |t|)  + c_1,
\end{equation*} 
again for appropriate constants \cite{Burrell}.
This means that due to quenched disorder,
the growth of entanglement is merely logarithmic in time, not linear.
There is an intuitive explanation for this:
The linear sound cone provided by the Lieb-Robinson bounds is replaced by
a logarithmically growing or even a constant one, leading to a suppressed 
entanglement propagation. A similar behavior is observed under
time-dependent fluctuating disorder \cite{Burrell2}.

\subsection{Topological entanglement entropy}

The topological entanglement entropy is a quantity that is 
constructed in a fashion that enables it to characterize quantum 
many-body states that exhibit {\it topological order}, a concept 
introduced in refs.\ \cite{OldWen,OldWen2} (see also 
refs.\ \cite{WittenOld,WenAdded}). On both sides of a critical point in a 
system undergoing a quantum phase transition, the quantum
many-body system may have a different kind of quantum order; but
this order is not necessarily one that is characterized by a
local order parameter: In systems of, say, two spatial
dimensions, topological order may occur. Topological order
manifests itself in a degeneracy of the ground state manifold 
that depends on the topology of the entire system and the 
quasi-particle excitations then show an exotic type of anyonic 
quasi-particle statistics. These are features that make
topologically ordered systems interesting for 
quantum computation, when exactly this degeneracy
can be exploited in order to achieve a quantum memory
robust against local fluctuations. They even allow in theory
for robust instances of quantum computation, then referred to
as {\it topological quantum computation} \cite{Kitaev,Freedman}.

The topological entanglement entropy is now designed as an
instrument to detect such topological order. Introduced in 
refs.\ \cite{Preskill,Wen}, it received significant attention 
recently \cite{FradkinTopological,Haque,Haldane,HammaLidar,Aguado,Kargarian,TrimerTopo}.
The details of the relationship between positive topological 
entanglement entropy and topological quantum order are discussed in
ref.\ \cite{Ortiz}. 

In ref.\ \cite{Preskill} a disc in the plane $I$ is considered with 
boundary length $L$. This disk is thought to be much larger than the 
correlation length, and it is hence assumed that an ``area law'' 
in the above sense holds. The entanglement entropy of $\rho_I$
will then have the form
\begin{equation}\label{topo}
        S(\rho_I) = \alpha L - \gamma + O(1),
\end{equation}
where the last term vanishes in the limit $L\rightarrow\infty$.
The prefactor $\alpha$ is non-universal and ultraviolet
divergent. However, $\gamma>0$ is an additive constant
which is universal and characterizes a global feature
of the entanglement in the ground state. This quantity
is referred to as {\it topological entanglement entropy} in 
ref.\ \cite{Preskill}.
To avoid ambiguities when distinguishing the constant term
from the linear one in eq. (\ref{topo}), ref.\ \cite{Preskill} 
makes use of the following construction: The plane is divided 
into four regions,
each of them being large compared to the correlation length.
$A$, $B$ and $C$ are arranged as neighboring each other
in three identical subparts of a disk. $D$ is the exterior
of the disk.  The respective reductions to the parts
are denoted as $\rho_A$ and $\rho_{AB}$ to
regions $A$ and jointly $A$ and $B$, 
respectively. The {\it topological entropy} $S_{\text{Topo}}$
is then defined as
\begin{eqnarray}\label{TopoP}
        S_{\text{Topo}} &=& S(\rho_A) + 
        S(\rho_B)+S(\rho_C)\\
        &-& S(\rho_{AB}) - S(\rho_{BC})
        -S(\rho_{AC}) + S(\rho_{ABC}).\nonumber
\end{eqnarray}
This is a linear combination of entropies of reductions, 
constructed specifically in a way such that the dependencies 
on the length of the respective boundaries of regions cancel.
It is not directly meant as an information theoretical quantity,
although the differences of entropies resembling a mutual information
expression. Also, slightly different definitions with similar 
properties are conceivable, and indeed, the independent
proposal of ref.\ \cite{Wen}  makes use of an alternative 
combination of entropies. The important aspect here is the 
above mentioned cancellation of the boundary term. Taking 
the behavior as in eq.\ (\ref{topo}) for granted, one indeed 
finds
\begin{equation*}
        S_{\text{Topo}} = - \gamma.
\end{equation*}
From the way $S_{\text{Topo}} $ is constructed it is 
a topological invariant, and depends only on a universal quantity 
(unaltered under smooth deformations, as long as one stays away
from critical points), and on the fashion how the regions 
are located with
respect to each other, but not on their specific geometry
(again assuming that the correlation length
is much smaller the regions and does not matter). Interestingly,
topological order is hence a global property that is detected by the
entanglement entropy. This construction
can also readily be used in numerical studies. 
The explicit computation how the entanglement entropy detects
the presence of topological order in an actually time-dependent model
undergoing a quantum phase transition from a spin-polarized 
to a topologically ordered phase 
has been systematically explored in ref.\ \cite{HammaLidar}, further
strengthening the findings of ref.\ \cite{Preskill}.

Since its proposal, this and related quantities have been
considered in a number of contexts. A natural candidate
to explore this concept is the {\it toric code state} of ref.\ \cite{Kitaev}:
Consider for this a square lattice $I=\{1,\dots, n\}^{\times 2}$ with
periodic boundary conditions, and place the physical
two-dimensional quantum spins on the vertices of this 
lattice.\footnote{Equivalently, 
one can place the physical spins on the edges and formulate the 
operators $\{A_s\}$ and $\{B_p\}$ as being 
non-trivially supported on the respective four
spins associated with vertices and plaquettes.}
This lattice is tiled into two sublattices of different color, red and white.
Every white $p$ and red plaquette $s$ 
is then associated with one of the commuting operators
\begin{equation}\label{ktcs}
        A_s=\prod_{j\in \partial s} \sigma_j^z,\,\,
        B_p=\prod_{j\in \partial p} \sigma_j^x,
\end{equation}
respectively, with non-trivial support on four spins each,
where as before 
$\sigma^x_i, \sigma^y_i, \sigma^z_i$ denote the Pauli operators supported
on $i$. The Hamiltonian of the system---a local Hamiltonian---is then taken to be
\begin{equation*}
        H=-\sum_s A_s -  \sum_p B_p.
\end{equation*} 
This is a gapped and frustration-free Hamiltonian. It is also straightforward
to verify that for any closed path $g$ the operator $\prod_{j\in g} \sigma_j^z$
commutes with all operators in eq.\ (\ref{ktcs}).
The ground state manifold depends on the topology of the lattice and is
in the chosen case four-fold degenerate. The topological entanglement
entropy, evaluated for this toric code state, gives $\gamma=\ln(2)$.
The ground states can readily be cast into a PEPS language, as has been
done in ref.\ \cite{PEPS}. An analysis how topological order 
can be grasped in a language of  entanglement renormalization or MERA
has been performed in ref.\ \cite{Aguado}: Indeed, the topological 
degrees of freedom can then be distilled to the top of the tensor network.

An equally important explicit and closely related 
model is the {\it loop model on a  
honeycomb lattice} of ref.\ \cite{Kitaev}. Ground states of 
more general string-net lattice models can also often
be expressed in terms of remarkably simple tensor 
networks \cite{Buerschaper,WenTensorNetworks,GuAdded}.
Entanglement entropies of {\it topological color codes} \cite{Bombin} 
have been studied in ref.\ \cite{Kargarian}. 
Equivalents of the topological entanglement entropy
for finite temperature---where the very robustness can be probed---have
been considered and introduced in refs.\ \cite{Castelnovo,PachosFiniteTemp}:
Notably, for Gibbs states it still makes sense to consider quantities of the 
type as in eq.\ (\ref{TopoP}), only with the respective entropies
being replaced by mutual informations grasping correlations instead of
entanglement, as discussed in detail 
in Section \ref{SectionCorrelations}. It is found that the interplay between thermal
effects, topological order and the size of the lattice indeed give rise
to well-defined scaling relations.

The study of entanglement entropies in fractional quantum
Hall states in a spherical geometry 
has been initiated in ref.\ \cite{Haque}; in ref.\
\cite{Haque2} Abelian Laughlin 
states as well as {\it Moore-Read states} have
been considered, where also rigorous upper bounds for {\it
particle entanglement entropies} have been derived. 
Particle partitioning entanglement in itinerant many-particle systems
has been studied in ref.\ \cite{Itinerant}. 
The MPS representation of the Laughlin wave function has been
derived in ref.\ \cite{Iblisdir}. The tolopogical entanglement of integer
quantum Hall states has been computed in ref.\ \cite{SierraTopo}.
Topological entanglement Renyi entropies have been considered in ref.\
\cite{FlammiaTopo}.
Similar quantities in {\it Chern-Simons theories}---the best understood
topological field theories---have been identified in ref.\ \cite{Dong}.
The suggestion that the full spectrum of $H$ in $\rho_I=\me^{-H}$
should be considered to detect topological order has been
proposed in ref.\ \cite{Haldane}. As being 
certified by this list of recent 
developments, studies of entanglement entropies as indicators
of topological order are still under rapid development.

\subsection{Relationship to black hole entropy}

As mentioned before, one of the particularly intriguing motivations 
for the study of area laws of the entanglement entropy is the 
suspected relationship to the area-dependence of the black hole 
entropy. The {\it Bekenstein-Hawking area law} 
\cite{Bardeen,Bekenstein,Hawking74} suggests that black hole 
carries an entropy that is proportional to its {\it horizon area} 
${\cal A}$,
\begin{equation*}
        S_{\text{BH}}=
        \frac{k c^3 {\cal A}}{4G\hbar}.
\end{equation*}
Hence, according to this
relationship, the (thermodynamical) entropy of a black hole is 
just a quarter of its area measured in Planck units \cite{Holographic},
i.e., when $k=c=G=\hbar=1$. For the 
sum of this black hole entropy and the matter entropy 
$S_{\text{Matter}}$ a second law of thermodynamics 
is proposed to hold. Such a generalized second law of thermodynamics 
led to the suggestion that one would have a ``spherical entropy 
bound'' for matter: In asymptotically flat spacetime, any weakly 
gravitating matter system would satisfy 
$S_{\text{Matter}}\leq 2\pi k E r/(\hbar c)$,
interestingly not containing $G$. $E$ denotes the total mass 
energy of the system, whereas $r$ stands for the smallest radius
of a sphere that contains the matter system at hand. The range 
in which one can expect the validity of such a law is discussed 
in ref.\ \cite{Holographic}.

The linear relationship between the boundary area and the 
(thermodynamical) entropy---formally, the two equations look 
identical ---suggests that one may expect a close relationship 
between these area laws: On the on hand, for the (von-Neumann) 
entanglement entropy of a subregion of a free quantum field in 
flat space time, on the other hand for the black hole entropy. 
This intriguing connection was first suggested and explored in
refs.\ \cite{Bombelli, Srednicki} and extended in refs.\
\cite{Callan,Larsen,PreskillOld}. Indeed, there are physical 
arguments that make the reduction of the situation of having 
a scalar field in a static spherically symmetric space-time 
to a scalar field in flat space time plausible \cite{Das}.
The exact status of the relationship between these quantities 
(or to what extent they are related by originating from a common 
cause---the general locality of interactions) is still subject 
to debate.\footnote{For a short general review on this connection 
see e.g., ref.\ \cite{Das}, for a calculation of the one-loop
correction to the Bekenstein-Hawking entropy in the presence 
of matter fields and its relationship to the geometric entropy 
see ref.\ \cite{Susskind}, and for an entanglement-based view 
of the Bekenstein-Hawking entropy law see refs.\ \cite{Brustein,Japanese}.}

This relationship has even been employed to take steps in computing 
the entanglement entropy in higher-dimensional conformal field 
theories: The AdS/CFT correspondence---relating a $d+2$-dimensional 
anti de Sitter (AdS) space to a $d+1$-dimensional conformal field theory (CFT)
\cite{Witten,Aharony}---has been made use of to study the Bekenstein
formula in the AdS context \cite{Japanese,Japanese2}, see also ref.\
\cite{Casini}. In this way, the above formula is used as a tool 
to compute the geometric entropy in a plausible fashion in 
situations where the exact computation is not known to be possible
using the tools of conformal field theory. 

The {\it holographic principle}---dating back to work in refs.\ 
\cite{tHooft,SusskindHolographic}---goes even further, and suggests 
that generally, all information that is contained in a volume of 
space can be represented by information that resides on the boundary 
of that region. For an extensive review, see ref.\ \cite{Holographic}.

\section{Area laws for classical systems and for total correlations}\label{SectionCorrelations}

\subsection{Classical harmonic systems}

Throughout this article, we have been concerned with
quantum systems on a lattice. What if we have classical
systems on a lattice, could one still expect an area law 
to hold?
Obviously, the concept of entanglement is no longer meaningful.
Also, the Shannon entropy of, say, the marginal distribution
of a distinguished region $I$ would not quantify correlations
in a reasonable fashion. What is worse,
in case of harmonic classical systems on a lattice, 
when thinking in terms of phase space cells, this quantity
is burdened with the usual Gibbs paradox.
However, it does make perfect sense to talk about {\it 
classical correlations in classical systems}, the 
appropriate quantity grasping such correlations 
being the mutual information:

Given a probability distribution $p$ on the lattice $L$,
one can quantify the correlations between the marginals 
with respect to a distinguished region $I$ and its complement $O$
by means of the {\it mutual information}. It tells us
how much information can be obtained on $O$ from 
measurements in $I$, and equally on $I$ by measurements in
$O$. This quantity enjoyes a number of very natural
properties. The mutual information is always positive---there can be
no negative correlations---and will vanish exactly if the 
probability distribution factorizes, in which case one can not learn
anything about $O$ from $I$. Given the marginals
$p_I$ and $p_O$ of the probability distribution $p$
on $I$ and $O$, respectively, the mutual information
is defined as
\begin{equation}\label{CMutual}
        I(I:O)= S(p_I) + S(p_O) - S(p),
\end{equation}
where here $S(p) = - \sum_{j} p_j \log_2 (p_j)$ is the standard
information theoretical {\it Shannon entropy}. It is noteworthy
that the mutual information does not suffer from the Gibbs 
paradoxon as will be shown below. How does the mutual information scale with the size 
of  a region in case of a harmonic coupled classical system? 
The subsequent statement clarifies this situation: Consider a 
{\it classical harmonic lattice system},
with Hamiltonian
\begin{equation}\label{HarmonicHamiltonian}
        H=\frac{1}{2}
        \biggl(
        \sum_{j\in L} p_j^2 + \sum_{j,k\in L}x_j V_{j,k} x_k
        \biggr),
\end{equation}
where now $x=(x_1,\dots, x_N)$ and
$p=(p_1,\dots, p_N)$ are the vectors of classical
position and momentum variables
of classical oscillators arranged on a cubic lattice
$L=\{1,\dots, N\}^{\times {\cal D}}$. 
The phase space coordinates are then
$\xi=(x,p)$. The matrix $V\in \rr^{|L|\times |L|}$
with a finite-ranged interaction defines the interaction.

The state of the system is 
defined by the phase space density, so a {\it classical 
distribution} $\rho:\rr^{N^{\cal D}}\rightarrow \rr^+$.
For any non-zero inverse temperature $\beta>0$,
this phase space distribution is nothing but
\begin{equation*}
        \rho_\beta(\xi) =\frac{1}{Z}e^{-\beta H(\xi)},\,\,\,
        Z= \int d\xi e^{-\beta H(\xi)}.
\end{equation*}
To define the mutual information, 
following the standard procedure, we split the phase
space into cubic cells each with a volume $h^{2 N^{\cal D}}$,
with $h>0$ being some constant. From the phase space
space density, we can then identify a discrete probability
distribution, from an 
average of the phase space density over these cells, 
$p_j = \int_{\text{Cell}} d\xi \rho(\xi)$ for $j\in L$. 
The discrete classical entropy is then defined as the
Shannon entropy of this probability distribution as
\begin{equation*}
        S_C(h) = - 
        \sum_{j\in L}
        p_j \log_2 (p_j).
\end{equation*}
We now come back to the situation of having a lattice system with 
an interior $I$ and an exterior $O$. The respective discrete 
classical entropies are defined as $S_I(h)$ and $S_O(h)$.
Obviously, the values of these entropies will depend on the choice 
of $h$, and in the limit $h\rightarrow 0$, they will diverge, 
logarithmically in $h$. This is a familiar observation in classical 
statistical physics, the divergence being resolved in the third 
law of thermodynamics. Here, we are, however, interested in 
classical correlations, as being quantified in terms of the 
mutual information which in the limit of $h\rightarrow 0$ is 
well-defined. Hence we can define the {\it classical mutual 
information of a harmonic lattice system} as
$I(I:O)= \lim_{h\rightarrow 0}
        \left(
        S_I(h) + S_O(h) - S_C(h)
        \right)$.
We are now in the position to state the area theorem for
classical harmonic systems \cite{Area3}:

\begin{theorem}[Correlations in classical harmonic systems]
Consider a harmonic lattice system 
with Hamiltonian as in eq.\ (\ref{HarmonicHamiltonian})
on a general lattice $G=(L,E)$. 
Then the classical mutual information $I(I:O)$ of the Gibbs
state at some inverse temperature $\beta>0$
satisfies an area law,
\begin{equation*}
        I(I:O)= O(s(I)).
\end{equation*}
\end{theorem}

The interesting aspect of this proof \cite{Area3} is that it 
relates this question of the {\it classical} mutual information to 
a quantity that arises in the quantum case in case where
the coupling matrix $V_x$ is replaced by $V_x^2$, and is
hence a simple corollary of earlier results 
on {\it quantum systems}, now with a coupling that is replaced by the squared
coupling matrix. Hence, a ``{\it quantum proof}''
can be applied to establish a statement on classical
lattice systems.  
The lesson to learn is that whenever one has local 
interactions---even in classical systems---one should not be
too surprised if this manifests itself in an area law in the
correlations. 

\subsection{Classical correlations quantum spin models}
\label{MutualInfo}
The situation is even simpler for finite-dimensional constituents. 
Indeed, quite in contrast to the overburdening technicalities that 
render the question of area laws in higher-dimensional quantum 
systems at zero temperature so difficult, the situation can here 
be clarified with hardly any mathematics at all: An elegant, 
but simple argument shows that total correlations in quantum 
(and classical) systems at non-zero temperatures always satisfy 
an area law. This is a statement on correlations---not entanglement, 
in contrast to the discussion of Subsection \ref{AreaThermal}--- in 
thermal states $\rho_\beta =e^{-\beta H}/Z$ for some $\beta>0$
for classical or quantum systems \cite{Correlations}. 
The relevant quantity grasping 
correlations is again the mutual information
\begin{equation}\label{QMutual}
        I(I:O)= S(\rho_I) + S(\rho_O) - S(\rho),
\end{equation}
where $S$ stands either for the von-Neumann quantum entropy, or for 
the Shannon entropy of the probability distribution. The classical 
variant was first discussed in ref.\ \cite{Area3}, the quantum 
version in refs.\ \cite{CasiniOld,Casini}. Ref.\  \cite{CasiniOld} 
introduces this quantity to avoid divergencies of the entanglement 
entropy in quantum field theory: In a similar fashion as above, 
regulators will in fact cancel each other, and the familiar 
{\it ultraviolet divergence in the quantum field limit} 
disappears. 

Interestingly, 
a general statement on the scaling of correlations at non-zero
temperature in terms of eq.\ (\ref{QMutual}) can be derived 
which holds for any spin model with local
dimension $d$ (see page 295 of ref.\ \cite{BratelliRobinson}
and the subsequent ref.\ \cite{Correlations}): 

\begin{theorem}[Classical correlations
at non-zero temperature]
Consider a classical or a quantum system with 
finite local dimension $d$ defined
on a translation-invariant lattice $G=(L,E)$. Consider the Gibbs
state at some inverse temperature $\beta>0$ of a local Hamiltonian $H$
with two-site interactions. In the classical case, where each of the lattice sites corresponds
to a spin with configuration space $\zz_d$,
\begin{equation}\label{CSimple}
        I(I:O)\leq |s(I)| \log (d).
\end{equation}
For a quantum system with local Hilbert spaces $\cc^d$, the mutual information 
satisfies the area law
\begin{equation}\label{QSimple}
        I(I:O)\leq \beta \|h\|  \, |s(I)|,
\end{equation}
where $\|h\|$ is the largest eigenvalue of all Hamiltonians
across the boundary of $I$ and $O$. 
\end{theorem}

This statement is valid in remarkable generality, 
given the simplicity of the argument. We will focus on 
quantum systems in the following. One can write the 
Hamiltonian $H$ having two-site interactions as
$       H= H_I + H_\partial + H_O$,
where $H_I$ and $H_O$ collect all interaction terms within the 
regions, whereas $H_\partial $ stands for terms 
connecting the two regions. The Gibbs state $\rho_\beta$ for
some inverse temperature $\beta>0$ minimizes the free energy
$F(\rho)= \text{tr}[H\rho] - S(\rho)/\beta$. Clearly, therefore,
\begin{equation*}
        F(\rho_\beta)\leq F(\rho_I\otimes \rho_O),
\end{equation*}
from which $I(I:O) \leq \beta \text{tr}[H_\partial (\rho_I\otimes \rho_O - \rho_\beta)]$
is obtained. As the right hand side depends only on terms
coupling the inside to the outside, i.e surface terms,  
eq.\ (\ref{QSimple}) follows straightforwardly.
A naive limit $\beta\rightarrow\infty$ will not yield an area law for zero temperature,
as the right hand side of eq.\ (\ref{QSimple}) then 
clearly diverges, but for any finite temperature, one obtains a
bound.

\section{Connection to simulatability}\label{Simulate}

There is an intimate connection between area
laws for the entanglement entropy and questions
of the simulatability of quantum many-body systems. 
The fact that there is ``little entanglement'' in a system
that satisfies an area law is at the core of the functioning 
of so powerful numerical techniques as the 
{\it density-matrix renormalization group} (DMRG) methods.
To describe the large research field of numerical 
simulation using DMRG-type methods would be
beyond the scope of the present review. Instead, we will
concentrate on the direct relationship between the 
``effective degrees of freedom'' that must be 
considered when classically describing quantum 
systems.

\subsection{Numerical simulations with the density-matrix renormalization group method}

This connection is particularly clear in one-dimensional systems, 
that is for quantum spin chains. Indeed, one can say that the fact 
that ground states of gapped systems satisfy an area law---and 
to a lesser extent that critical systems merely have a logarithmic 
divergence of the entanglement entropy---is responsible for the
success of the density-matrix renormalization approach. Matrix-product 
states also satisfy a one-dimensional area law. As MPS are
underlying the DMRG approach this suggest that the 
entanglement content of a state and the best possible 
performance of a DMRG approach can be intimately linked.

Historically, DMRG was born out of an idea of renormalization, where one 
iteratively identifies the relevant degrees of freedom, grasping the 
essential physics of the problem, when going from one step of the procedure to the 
next one. This general idea goes back to the {\it real-space
renormalization group approach}, presented  in 
ref.\ \cite{Wilson1975} in the mid 1970ies. This approach was particularly 
successful in the numerical assessment of the Kondo problem, whereas
for other problems, results were not quite what was hoped for.
The birth of the DMRG approach as such was related to a clear analysis
of the strengths and weaknesses of the real-space renormalization group
approach to study the low-energy properties of quantum many-body 
systems \cite{WhiteNoack}. Ref.\ \cite{White} is seen as the 
manuscript in which the DMRG method has actually
been introduced. Since then, 
this method has seen a standard method in the numerical study of 
strongly correlated quantum many-body systems. For a recent
review, see ref.\ \cite{Scholl}.

Initially, the formulation of DMRG was based on the above renormalization
idea. However, in the following years it became clear that DMRG generates
matrix-product states, an insight that has been reported in ref.\ 
\cite{Rommer} for the thermodynamical limit of DMRG, and in ref.\ 
\cite{MPSDMRG} for finite-size DMRG methods with the latter placing
a particular emphasis on exploiting a rotational symmetry in variational 
approaches. Ref.\ \cite{PeschelDMRG} gives a relatively early exhaustive 
overview over variational ansatzes with matrix-product states and the 
relationship with the DMRG idea. Ref.\ \cite{Rommer} already hinted at 
the possibility for treating period boundary conditions in the MPS picture
but chose translation invariant matrices. Ref.\ \cite{Frank} relaxed
this constraint to demonstrate that a suitable formulation significantly 
outperforms standard DMRG for periodic boundary conditions in terms
of memory requirements.

Hence, DMRG---in its several variants---can be seen as a variational 
method, where the optimization problem 
\begin{eqnarray}
        \text{minimize} & \langle \psi|H|\psi\rangle,\\
        \text{subject to} & |\psi\rangle \in (\cc^d)^{\otimes N},
        \nonumber
\end{eqnarray}
impractical already because of its exponentially large feasible set,
is replaced by a variant of an optimization problem over a polynomially large set
\begin{eqnarray}
        \text{minimize} & \langle \psi |H|\psi\rangle,\label{MPSMin}\\
        \text{subject to} & |\psi \rangle\in (\cc^d)^{\otimes N}  \text{ is an MPS 
        vector of dimension $D$.}\nonumber
\end{eqnarray}
In this variant, or---more accurately---in each of these variants
one does not attempt in one go to identify the global optimum, but rather
effectively iteratively solves for the local matrices involved. Such an iteration 
will then certainly converge (albeit strictly speaking
not necessarily to the 
global minimum).\footnote{For mixed state simulations, see refs.\ \cite{Mixed,Animesh}.}

\subsection{Approximation of states with matrix-product states}

Any such method, can then only be as good as the best possible 
MPS can approximate the true ground state at hand. This, in fact, 
is related to the entanglement content, in that it matters whether 
or not the true ground state satisfies an area law or not. In the 
light of previous discussions, this connection is not that surprising 
any more:  After all MPS satisfy an area law for the entanglement 
entropy. Hence, one aims at approximating ground states with states
that have in this sense little entanglement, and those states
can be well approximated by MPS that satisfy an area law in the 
first place.

This connection has been hinted at already in the first work
on DMRG \cite{White}, where the spectrum of the half chain 
has been considered and put into relationship with
the {\it ``truncation error''} in DMRG. This is a key figure of merit of 
the quality of an approximation in a step, so unity minus
the weight of those terms being kept in a step of the iteration.

This connection between the decay of spectral values of half chains,
the more rapid the decay the better can DMRG perform, has been made 
more precise and fleshed out in ref.\ \cite{PeschelEntanglementDMRG}. 
In ref.\ \cite{Latorre2} the relationship to criticality in this 
context has been emphasized. Ref.\ \cite{LatorreReview} is a short 
review on this question. In more recent quantitative approaches, the 
optimal approximation that can possibly be obtained by a MPS of a 
given $D$ is considered. Let us denote with
${\cal H}_N= (\cc^d)^{\otimes N}$ the Hilbert space of a
quantum chain of length $N$. MPS are considered as defined
in eq.\ (\ref{MPS}) for open boundary conditions.
Given a family $\{|\psi_N\rangle\}_N$ of state vectors, it is said that it
{\it can be approximated efficiently
by MPS} if for every $\delta>0$ there exists a sequence 
$|\psi_{N,D(N)}\rangle $ of MPS with 
        $D(N) = O(\text{poly}_\delta (N))$
such that
\begin{equation*}
        \|\, |\psi_N\rangle\langle \psi_N| - |\psi_{N,D(N)}\rangle\langle\psi_{N,D(N)}|\, \|_1\leq \delta,
\end{equation*}
where $\|.\|_1$ denotes the usual trace-norm. In contrast, it is said that this sequence
{\it cannot be approximated efficiently by MPS} if there exists some $\delta>0$ such that 
no sequence of MPS with $D(N)$ growing 
polynomially can approximate $|\psi\rangle\langle \psi|$
up to a small error 
$\delta$ in trace-norm \cite{SchuchApprox}:

\begin{theorem}[Approximatability with MPS]
\label{Ap}
Consider sequences of state vectors $ \{|\psi_N\rangle\}_N\in {\cal H}_N$ 
of a quantum chain of length $N$, and denote as before the reduced state of a block 
$I=\{1,\dots, n\}$ of length $n$ with $\rho_I$. If the sequence 
of $\rho_I$ satisfies an area law for a Renyi entropy $S_\alpha$ 
for $\alpha<1$,
\begin{equation*}
        S_\alpha(\rho_I)=O(1),
\end{equation*}
then the sequence $ \{|\psi_N\rangle\}_N$ is efficiently 
approximable by MPS. In contrast, if the von-Neumann entropy 
        $S_1(\rho_I)=\Omega(n)$, 
so grows at least linearly with the block size, then it cannot
be approximated efficiently by MPS. This means that states 
satisfying a volume law cannot be approximated. The same holds 
true if any Renyi entropy $S_\alpha$ for some $\alpha>1$ grows 
at least as $S_\alpha(\rho_I)=\Omega(n^\kappa)$ for some $\kappa<1$. 
Otherwise, the connection is undetermined, in that examples
both for approximable and inapproximable states can be found.
\end{theorem}

This statement clarifies the connection between the entanglement content and the possibility
of describing states with matrix-product states. The validity of an area law implies that there
is sufficiently little entanglement in the state such that an economical description in terms
of matrix-product states is possible. The enormous success of DMRG is related to the fact that
gapped systems satisfy an area law. Even if the system is critical, the logarithmic divergence 
still allows for a relatively economical description in terms of matrix-product states. The fact
that Renyi-entropies for $\alpha$ smaller than or larger than unity feature here may be
seen rather as a technical detail. The general message is clear: The area-like entanglement 
scaling, with or without 
small corrections, allows for an efficient 
approximation in $D$ for matrix-product states.

To reiterate the point made in Subsection \ref{noneq}: Quenched, 
non-equilibrium systems can indeed fall exactly into the category 
of having an effectively linearly growing block entropy, so are 
characterized by a volume law for the entanglement entropy. More 
precisely, we face the interesting situation that while for each 
time, we have an area law in $n$, the constant in the upper bound 
grows in time such that for a suitable choice for the sub-block, 
one arrives effectively at a volume law, as made precise in 
Theorem \ref{QuenchTheorem}. This has severe 
practical implications: 
For small times, {\it t-DMRG} \cite{VidalTEBD,Daley,WhiteFeiguin,Kollath,QuenchDMRG,Scholl},
the {\it time-dependent version of DMRG}, can very accurately keep track 
of the dynamics of the system. This is a variant in which one essentially 
makes a Lie-Trotter approximation of the time evolution operator, and 
then approximates in each time step the resulting state vector by an 
MPS, going back to ref.\ \cite{VidalTEBD}. The functioning of this
algorithm can essentially be traced back to the observation that 
an arbitrarily good approximation to the propagator can be established
with polynomial computational resources in the system size \cite{MPS}. 
In time, however, one 
will eventually encounter typically an exponential increase in the 
number of degrees of freedom to be kept in order to faithfully describe 
the state. This eventually limits the time up to which one can 
numerically simulate time evolution using a variant of DMRG.
The increase in the entanglement content 
also eventually limits classical simulations of 
{\it quantum adiabatic algorithms} 
based on MPS, which nevertheless perform often 
impressively well (for a careful numerical analysis, see, e.g.,
ref.\ \cite{AdiabaticComp}). It is interesting to note, 
however, that this complexity does not
necessarily translate in the difficulty of following the time-evolution
of specific observables when evolving them in the 
{\it Heisenberg picture}
using t-DMRG. Then, in some cases the Heisenberg time evolution
can be carried out exactly for finite bond dimension and arbitrary 
long times \cite{HartmannP08,ProsenP07,ZnidaricPP08}.

There are numerical simulation methods that allow for the simulation 
of certain quantum states that do not satisfy an area law. MERA already 
allows for a logarithmic divergence of the entanglement entropy in 
one-dimensional systems. Weighted graphs state based approaches
\cite{Weighted} and its 1-D variant, the {\it renormalization algorithm
with graph enhancement} \cite{Rage} can cope with instances of volume 
laws for the entanglement entropy, the latter in 1-D the former in 
arbitrary spatial dimensions. Early work on the simulation of a particular 
kind of discrete time evolution, namely the application of random unitary 
circuits, suggests that this may be a promising approach for the 
``efficient simulation of quantum many-body systems beyond area laws''.

We end this subsection with a note rather from the computer science than
from the physics perspective: The fact that a true ground state is well-approximated by an 
MPS does, strictly speaking, 
not necessarily mean that DMRG will also efficiently {\it find} this best approximation.
In practice, DMRG works well, and it typically produces good and
reasonable results. It is remarkable how well this approximation is
found in the iterative scheme as being pursued by any DMRG algorithm:
After all, the full problem eq.\ (\ref{MPSMin}) is a non-convex polynomial 
global optimization problem of very high order ($\langle \psi|H|\psi\rangle$ is 
of degree $N^2$ in $D$). Still, by local variations and sweeping one achieves very 
good results. The ultimate reason for this impressive performance is yet
to be ultimately understood.

Having said that, the worst case complexity of the problem of 
finding the best approximation can be computationally difficult  
in the sense of computer science. In fact, the class of problem 
of keeping some matrices fixed and varying over a finite subset 
has in worst case instances that are {\it NP-hard} \cite{NP}. In 
non-translation invariant settings, one even finds that if 
one could efficiently identify the best possible MPS approximation, 
one could solve efficiently NP-hard problems \cite{SchuchNP}.
Even more strongly put, the 
problem of approximating the ground state energy of a system 
composed of a chain of quantum systems is 
{\it QMA-complete} \cite{GottesmanQMA}.

This should be seen as a warning sign: The functioning of
variational algorithms such as DMRG is essentially based on 
heuristics, and in worst case  one can encounter hard problems. 
The energy landscape is then so rugged that one gets stuck in 
local optima. Still, while it is important to acknowledge that 
DMRG is strictly speaking not certifiable, it is still true 
that it works very well in practice and is one of the pillars 
of the numerical assessment of strongly correlated systems
in 1-D.

\subsection{Implications on higher-dimensional simulations}

For higher dimensional systems, tensor-product states or PEPS, as 
well as those of MERA, satisfy area laws, as has been discussed 
in Subsection \ref{PEPS}. This fact suggests that when minimizing 
$\langle \psi |H|\psi\rangle$ for an $N\times N$-lattice subject 
to $|\psi \rangle\in (\cc^d)^{\otimes N^2}$ being a PEPS or MERA 
described by polynomially many real parameters, one encounters a 
good approximation whenever the system at hand already satisfies 
an area law. In the light of the fact that even critical 
two-dimensional systems can satisfy an area law, this would mean 
that they can be well-described by PEPS or MERA described by 
relatively few parameters. Numerical work in case of PEPS 
indicates that this is indeed the case \cite{PEPSOld,2DHardCoreBosons,2DPEPS,VerstraeteBig}, 
even in the thermodynamical limit of an infinite system \cite{iPEPS} or for
fermions \cite{Zhou,SchuchFermiPEPS,MERAF3}.
 
A rigorous result similar to Theorem \ref{Ap}, yet, is still 
lacking for PEPS or MERA. The  intuition developed so far, 
however, is in one way or the other quite certainly right: Whenever 
an area law is satisfied, PEPS with small bond dimension should 
give rise to a reasonably good approximation. Here, subtle 
aspects are rather connected to the fact that the exact contraction 
of the tensor networks of PEPS, and hence the computation of
expectation values, is inefficient, and that approximate contractions 
have to be employed. Suitable subsets, such as the class of 
{\it string states}, can always be efficiently contracted, 
giving rise to very promising variational sets in higher-dimensional 
systems \cite{Strings}. The method in ref.\ \cite{NonMPS} 
also gives rise to 
certifiable approximations of 2-D ground states for a class of
models, exploiting quasi-adiabatic evolutions.

As before, one has to distinguish the variational set as such from the computational
method of varying over this set. Usually, one has to find practical and heuristically 
suitable methods of solving a
global optimization problem over many variables. Several strategies may be followed when
varying over suitable sets to simulate higher-dimensional strongly correlated systems:
One may use {\it local variations such as in DMRG}, 
{\it imaginary time evolution}, or {\it flow methods} \cite{Flow}, 
making use of gradient flow and 
optimal control ideas to vary over the manifold
of unitary gates that describe the variational set of states at hand. For MERA, 
the same intuition should hold true. Here, first approaches implemented
have been focused on one-dimensional systems \cite{Flow,Rizzi,MERA2}, but the 
ideas are also applicable in 
higher dimensions \cite{EvenblyNew,MERAF2,MERAF3}.
  
\section{Perspectives}

In this Colloquium, we have presented the state of affairs in
the study of area laws for entanglement entropies. 
As has been pointed out above, this research field is 
presently enjoying a lot of attention, for a number 
of reasons and motivations. Yet, needless to say, there are
numerous open questions that are to be studied, of which 
we mention a few to highlight further perspectives:
\begin{itemize}
\item Can one prove that gapped higher-dimensional
general local lattice models always satisfy an area law?
\item In higher dimensional systems, critical systems can both
satisfy and violate an area law. What are further conditions to
ensure that critical systems satisfy an area law? What is the
exact role of the Fermi surface in the study of area laws in
fermionic critical models?
\item Can one compute scaling laws for the 
mutual information for quasi-free systems?
\item For what 1-D models beyond quasi-free and conformal
settings can one find rigorous expressions for the
entanglement entropy?
\item Under what precise conditions do quenched 
disordered local models 
lead to having ``less entanglement''?
\item What are the further perspectives of using conformal 
methods for systems with more than one spatial dimension?
\item Can the link between the Bekenstein formula in the
AdS context and the scaling of geometric entropies in
conformal field theories be sharpened?
\item To what extent is having a positive topological entropy
and encountering topological order one to one?
\item How can the relationship between satisfying an area law and 
the efficient approximation of ground states with 
PEPS be rigorously established?
\item What efficiently describable states satisfy an area law,
such that one can still efficiently compute local properties?
\item Are there further instances for 1-D systems
satisfying an area law that allow for certifiable
approximations in terms of matrix-product states?
\end{itemize}
These questions only touch upon the various perspectives that open
up in this context. The quantitative study of a research area
that could be
called ``Hamiltonian complexity''\footnote{This term has been coined by
B.M.\ Terhal.} is just beginning to 
emerge. The puzzle of 
how complex quantum many-body systems are, and how
many effective degrees of freedom are 
exploited by nature, is still one of the intriguing topics in
the study of interacting quantum systems.

\section{Acknowledgements}

We acknowledge very fruitful and stimulating
discussions on the subject of this paper with a number
of friends and colleagues over the last years. Among them, we specifically mention
S. Anders,
K.\ Audenaert,
H.J.\ Briegel,
P.\ Calabrese,
H.\ Casini,
J.I.\ Cirac, 
C.M.\ Dawson,
M.\ Fannes,
S.\ Farkas,
R.\ Fazio,
A.\ Fring,
D.\ Gross,
M.J.\ Hartmann,
M.B.\ Hastings,
J.\ Keating,
V.\ Korepin,
J.I.\ Latorre,
R.\ Orus,
T.J.\ Osborne,
J.\ Pachos,
I. Peschel,
J.\ Preskill,
U.\ Schollw{\"o}ck,
N.\ Schuch,
F.\ Verstraete,
G.\ Vidal,
J.\ Vidal,
R.F.\ Werner,
M.M.\ Wolf, and P.\ Zanardi.  
We specifically thank
T.\ Barthel,
P.\ Calabrese, 
D.\ Gottesman,
M.B.\ Hastings, 
J.I.\ Latorre, 
T.J.\ Osborne,
I.\ Peschel, and
X.-G.\ Wen
for valuable feedback on a draft version of this review.
This work has been supported by the DFG (SPP 1116), the EU 
(QAP, COMPAS, CORNER, HIP), the 
Royal Society, the EPSRC QIP-IRC, 
Microsoft Research, and the EURYI Award Scheme.

\section{Appendix: Fisher-Hartwig Theorem}
In this appendix we briefly present an important
technical result concerning the asymptotic behavior 
of Toeplitz matrices \cite{BoettcherS06}.

\begin{lemma}[Fisher-Hartwig]
        Consider a sequence of $n\times n$ Toeplitz matrices $\{T_n\}_n$ with entries $(T_n)_{i,j}=(T_n)_{i-j}$,
        \begin{equation*}
                (T_n)_{l} = 
                \frac{1}{2\pi}\int_0^{2\pi}
                \md\phi\, g(\phi) \me^{-\mi l \phi},
        \end{equation*}
 generated by $g:[0,2\pi)\rightarrow\cc$. 
        Let $g$ be of the form
        \begin{equation*}
                g(\phi)= b(\phi) \prod_{r=1}^R t_{\beta_r}(\phi-\phi_r)
                u_{\alpha_r}(\phi-\phi_r),
        \end{equation*}
        with
        $t_\beta(\phi) = \me^{-\mi\beta (\pi - \phi)}$,
                $u_\alpha = (2-2\cos(\phi))^\alpha$,
                 $\text{Re}(\alpha)>-1/2$, and
        $b:[0,2\pi)\rightarrow\cc$ 
         a smooth non-vanishing function with winding number zero.
        Then \cite{Boettcher,Libby,Basor}, 
        for $|\text{Re}(\alpha_r)|<1/2$ and 
        $|\text{Re}(\beta_r)|<1/2$ or $R=1$, $\alpha=0$
        $|\text{Re}(\beta)|<5/2$, the asymptotic behavior of the
        determinant of $T_n$ is given by
        \begin{equation*}
                \lim_{n\rightarrow\infty}
                \frac{\text{det}(T_n) }{ E G^n n^{\sum_{r=1}^R(
                \alpha_r^2 - \beta_r^2)}}=1
        \end{equation*}
        where $E=O(1)$ in $n$
        and 
        \begin{equation*}
                G= \exp\left(
                \frac{1}{2\pi} \int_0^{2\pi} \md\phi\, \ln(b(\phi))
                \right).
        \end{equation*}
\end{lemma}


\end{document}